\documentclass{article} 
\usepackage{reler_preprint,times}
\renewcommand{\headrulewidth}{0.4pt}

\usepackage[hyperfootnotes=false]{hyperref}
\usepackage{url}
\hypersetup{hidelinks}
\usepackage{graphicx}
\usepackage{caption}
\usepackage{placeins}
\usepackage{float}
\usepackage{booktabs}
\usepackage{amsmath}
\usepackage{amsthm}
\usepackage{amsfonts}
\usepackage{amssymb}
\usepackage{microtype}
\microtypesetup{expansion=false}
\usepackage{xcolor}
\usepackage{colortbl}
\usepackage{pifont}
\usepackage{multirow}
\usepackage{algorithm}
\usepackage{algpseudocode}
\usepackage{mathtools}

\definecolor{relergreen}{HTML}{1BAF7A}
\colorlet{relerrow}{relergreen!10!white}
\DeclareMathSizes{7.5}{7}{5}{5}
\DeclareMathSizes{6.5}{6}{5}{5}
\newcommand{\tablesetup}{\fontsize{8}{9.5}\selectfont
  \renewcommand{\arraystretch}{1.08}\setlength{\tabcolsep}{4pt}}
\newcommand{\compacttablesetup}{\tablesetup}
\newcommand{\tablehead}[1]{\textbf{#1}}
\newcommand{\tablescore}[2]{#1\,{\fontsize{6.5}{7}\selectfont\ensuremath{\pm}\,#2}}

\newtheorem{proposition}{Proposition}

\newif\ifshowtodos
\showtodosfalse
\ifshowtodos
  \newcommand{\todonote}[1]{{\footnotesize\textcolor{red}{[\textbf{TODO:} #1]}}}
\else
  \newcommand{\todonote}[1]{}
\fi

\definecolor{taxroot}{HTML}{264653}
\definecolor{taxacq}{HTML}{4F86C6}
\definecolor{taxdep}{HTML}{E9A23B}
\definecolor{taxevo}{HTML}{2A9D8F}
\definecolor{taxev}{HTML}{E76F51}
\definecolor{taxsec}{HTML}{8E5EA2}

\newcommand{\email}[1]{\href{mailto:#1}{\texttt{#1}}}

\title{\raggedright Learning to Retrieve via Reinforcement Learning in Embedding Space}

\author{Qi Liu, Fengming Liang, Yiqun Chen, Erhan Zhang, Jiaxin Mao\\
Renmin University of China\\
\email{qiliu6777@gmail.com}, \email{maojiaxin@gmail.com}
}

\iclrfinalcopy

\begin{document}

\maketitle

{\renewcommand{\thefootnote}{}\footnotetext{Code: \url{https://github.com/liuqi6777/RELER}}}

\begin{abstract}
Dense retrieval models are typically trained with contrastive objectives
that learn effective representations but do not directly optimize retrieval
metrics or downstream task performance. To address this problem, we introduce \textbf{RELER}
(\textbf{RE}inforcement \textbf{LE}arning for \textbf{R}etrieval), a reinforcement
learning framework that enables existing embedding models to \emph{learn to retrieve}
directly in embedding space and align to task-specific rewards.
We train RELER by sampling unit-length query and document embedding actions from
von Mises--Fisher (vMF)
distributions centered on normalized encoder outputs, scoring the resulting retrieval
or downstream outcomes as rewards, and updating the encoder with REINFORCE using a
leave-one-out baseline (RLOO).
As exploration in the high-dimensional embedding space is prone to sampling noise,
we further propose conditional-mean projection (CMP), which projects each sampled
embedding onto the low-dimensional subspace spanned by its encoder output and the
candidate embeddings it is compared against, reducing noise in the policy gradient
while preserving its expectation.
We evaluate RELER on BRIGHT, a benchmark
with reasoning-intensive queries that remain challenging for existing embedding
models. RELER consistently outperforms InfoNCE and LambdaLoss in average nDCG@10 when
post-training BGE-M3 and Qwen3-Embedding backbones.
We further evaluate downstream utility through retrieval-augmented generation (RAG),
where we adapt only the query encoder while keeping the document index and
generator fixed. Across seven QA datasets, jointly optimizing retrieval and
answer rewards improves both average retrieval performance and answer quality
in RAG.
\end{abstract}

\section{Introduction}

Dense retrieval models enable large-scale semantic retrieval by encoding
queries and documents into a shared embedding space, where relevance is
modeled by vector similarity
\citep{karpukhin2020dense,zhang2025qwen3embedding}.
They support applications including web search \citep{huang2013learning},
retrieval-augmented generation (RAG) \citep{lewis2020retrieval}, agentic search
\citep{jin2025searchr1}, and memory recall for long-horizon agents
\citep{packer2023memgpt}. Despite these varied uses, most text embedding models
are trained with large-scale contrastive learning,
encouraging relevant query--document pairs to have higher similarity than
irrelevant ones \citep{zhan2020learning,wang2022text}.
This training objective, however, is only indirectly related to retrieval
metrics such as nDCG or downstream task performance such as answer accuracy
in RAG. Reducing the contrastive loss therefore does not necessarily
translate into better retrieval quality or downstream utility
\citep{zhang2024utility}.

This objective mismatch is analogous to that in language model alignment:
next-token prediction does not directly optimize human preferences or task
success. Reinforcement learning (RL) provides a principled approach by
optimizing rewards derived from human preferences or task outcomes
\citep{ouyang2022training,deepseekai2025r1}.
This motivates using RL to align retrieval models directly with retrieval
and downstream metrics, as RL can optimize such non-differentiable objectives
by sampling retrieval decisions, evaluating their outcomes with the target
metrics, and estimating policy gradients without differentiating through the
retrieval process or the downstream evaluator
\citep{williams1992simple}.
Recent work has explored RL for embedding and retrieval models through
policies over generated text \citep{sun2026grace,jiang2026embedrl} or discrete
document selections \citep{salemi2024optimization,liu2025retrievalgrpo,zhang2026harr}.
These policies explore text or document choices; the query and document
vectors used for similarity scoring are not themselves sampled as actions.

In this work, we revisit reinforcement learning for dense retrieval by
treating embeddings as learnable retrieval actions. Because their
similarities determine which documents are retrieved, exploring alternative
embeddings allows task feedback to guide the representations themselves.
We introduce \textbf{RELER} (REinforcement LEarning for Retrieval), a
framework for post-training embedding models to \emph{learn to retrieve}:
sample embedding actions, evaluate the retrieval results or downstream answers they
produce, and update the encoder to favor actions with higher rewards.
The framework explores both query and document embeddings, and also
supports query-only adaptation against a fixed document index. Inference
retains deterministic single-vector retrieval.

To turn this view into a practical learning algorithm, RELER places a von
Mises--Fisher (vMF) policy around each normalized encoder output, enabling
exploration on the unit sphere used for cosine retrieval
\citep{bendada2025exploring}. To optimize retrieval, RELER independently
samples query and document embeddings and uses product rollouts
\citep{kuntz2022product} to pair every sampled query with every document
bundle. Averaging rewards across these pairings gives
each action feedback from multiple retrieval outcomes without additional
encoder passes. REINFORCE with a leave-one-out baseline
\citep{kool2019buy} then converts this feedback into updates to the shared
encoder. High-dimensional exploration, however, also introduces random
components that leave retrieval scores unchanged but add gradient noise.
Building on document-space projection \citep{wang2019variance}, we develop
\emph{conditional-mean projection} (CMP): for each reward term, sampled
actions are projected onto the subspace spanned by their policy mean and
the directions that determine the reward. Applied only in gradient
computation, CMP averages out this irrelevant randomness while preserving
the relevant scores and the expected policy gradient.

We evaluate RELER as a post-training method for existing embedding models,
first on reasoning retrieval and then on downstream RAG tasks.
Existing embeddings receive extensive training on tasks represented
in MTEB \citep{muennighoff2023mteb} and already perform strongly
\citep{chen2024bgem3,zhang2025qwen3embedding}. We therefore focus on BRIGHT
\citep{su2025bright}, whose reasoning-intensive queries expose remaining
weaknesses, to test whether limited additional supervision can improve these
models. Using fewer than 5k queries, RELER consistently outperforms supervised
baselines trained on the same data across BGE-M3 and Qwen3-Embedding backbones.
For downstream RAG, we test whether task feedback improves answer quality by
adapting only the query encoder with the index and generator frozen.
Combining retrieval and answer rewards improves average retrieval and answer
quality over the pretrained encoder across seven QA datasets.

\section{Related Work}
\label{sec:related-work}

\paragraph{Training dense retrievers and embedding models.}
Contrastive learning dominates dense retrieval: dual encoders learn a shared
geometry by scoring relevant query--document pairs above negatives
\citep{karpukhin2020dense}, while hard-negative mining and improved sampling
sharpen these comparisons \citep{xiong2021approximate,qu2021rocketqa}.
General-purpose embedding models scale this recipe with weak supervision,
multi-stage and multi-task training, and multilingual or multi-function objectives
\citep{wang2022text,li2023towards,xiao2024cpack,chen2024bgem3}. LLM-based
models adapt decoder-only backbones through bidirectional attention or
specialized pooling, together with synthetic pairs, instructions, and
contrastive fine-tuning
\citep{behnamghader2024llm2vec,wang2024improving,lee2025nvembed,
muennighoff2025generative,zhang2025qwen3embedding}. These methods provide the
pretrained dual encoder that RELER subsequently adapts with task-level rewards.
ReasonIR, RaDeR, and DIVER develop retrievers for reasoning-intensive queries,
while ReasonEmbed uses reasoning-aware synthetic data and adaptive contrastive
training \citep{shao2025reasonir,das2025rader,long2025diver,chen2026reasonembed}.

\paragraph{Reward-based adaptation of retrievers and embedding models.}
Existing RL approaches differ in the action they sample and how rewards reach
the encoder. ROPG-RL samples a personal document from a categorical policy and uses
downstream generation quality as reward \citep{salemi2024optimization}.
Retrieval-GRPO assigns multi-objective rewards to dynamically retrieved
products and optimizes the dense retriever through their similarity scores
\citep{liu2025retrievalgrpo}, whereas HARR samples ordered document lists from
a Plackett--Luce policy and uses answer-level rewards \citep{zhang2026harr}.
These policies reshape retrieval geometry through document selections or scores.
Another approach uses generated reasoning: GRACE pools rationale-conditioned states into
embeddings, while Embed-RL trains a reasoner using feedback from a separate
embedder \citep{sun2026grace,jiang2026embedrl}.
For large action sets, \citet{bendada2025exploring} sample a state embedding
from a vMF distribution and select actions through nearest-neighbor search.
RELER independently samples both query and document embeddings and uses retrieval rewards
with CMP to update a shared encoder. Deployment uses deterministic
single-vector encoder outputs.
Appendix~\ref{sec:extended-related-work} discusses additional retrieval
methods and the prior work underlying our policy-gradient estimator.

\section{Method}

\begin{figure}[t]
 \centering
 \includegraphics[width=\textwidth]{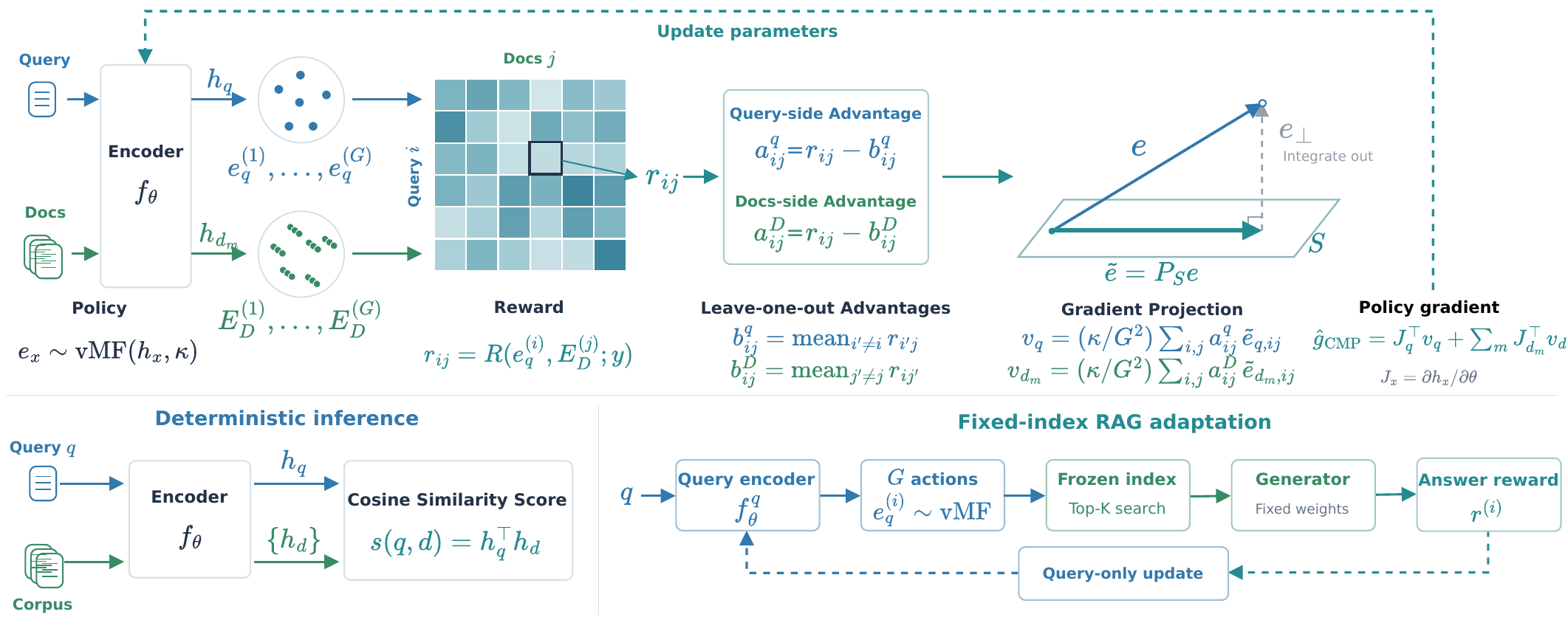}
 \caption{RELER workflow. Top: embedding rollouts yield rewards for RLOO+CMP
 updates. Bottom left: deterministic cosine retrieval at inference.
 Bottom right: query-only adaptation with a frozen document index and generator.}
 \label{fig:query-slate-method}
\end{figure}

We present the embedding-space policy, product-rollout RLOO, and
conditional-mean projection, then describe fixed-index query adaptation and
the training rewards, as illustrated in Figure~\ref{fig:query-slate-method}.

\subsection{Retrieval as an Embedding-Space Policy}
\label{sec:embedding-policy}

Let $h_x=f_\theta(x)\in\mathbb S^{d-1}$ be a normalized text representation.
The query and document roles share the trainable encoder.
At deployment, the similarity of query and document is defined by
$h_q^\top h_d$. Small changes in embedding direction can change the score,
therefore influencing retrieval outcomes or downstream task performance.
Exploring around the encoder outputs lets retrieval or answer feedback guide the
representations, even when the reward is not differentiable.
To explore on the unit sphere, we let each representation define a
von Mises--Fisher (vMF) policy during training
\citep{mardia2000directional, bendada2025exploring}:
\begin{equation}
 \pi_\theta(e_x\mid x)=C_d(\kappa)\exp(\kappa h_x^\top e_x),
 \qquad e_x\in\mathbb S^{d-1}.
 \label{eq:vmf-policy}
\end{equation}
Here $C_d$ normalizes the density and $\kappa>0$ controls exploration.
We specify $\kappa$ through the expected alignment
$\rho=\mathbb E[h_x^\top e_x]$; larger $\rho$ concentrates actions around
the deterministic representation. We describe the detailed sampling procedure
in Appendix~\ref{sec:sampling-settings}.

A training record $(q,D,y)$ contains a query, valid candidates
$D=(d_1,\ldots,d_n)$, and task supervision $y$, such as relevance labels
for retrieval or reference answers for RAG.
The policy $\pi_\theta(\cdot\mid q,D)$ samples the query action $e_q$
and each document action in $E_D=(e_{d_1},\ldots,e_{d_n})$ independently.
The training objective is
\begin{equation}
 \mathcal{J}_\kappa(\theta\mid q,D,y)
 =\mathbb E_{\pi_\theta(\cdot\mid q,D)}
    [R(e_q,E_D;y)].
 \label{eq:expected-reward-objective}
\end{equation}
Here $R$ is a task-defined scalar reward that evaluates the retrieved document list
or a downstream output against $y$. It need not be differentiable, so the
same objective accommodates both retrieval metrics and downstream task
metrics.

\subsection{Product Rollouts with REINFORCE-LOO}
\label{sec:policy-objective}

For each training record, we sample a rollout group consisting of
$G\geq2$ query actions and $G$ document bundles.
One-to-one pairing would give each query action feedback from only one
document bundle, making its evaluation depend on that particular draw.
Product rollouts \citep{kuntz2022product} evaluate all pairings within the
group, so every query action is compared against the same document bundles
and every bundle against the same query actions:
\begin{equation}
 r^{(i,j)}=R(e_q^{(i)},E_D^{(j)};y),
 \qquad i,j=1,\ldots,G.
 \label{eq:reward-matrix}
\end{equation}
Reusing actions yields $G^2$ correlated rewards without additional encoder
passes or action samples, at the cost of more score and reward evaluations.

To evaluate each action, we average over the opposite role's draws.
The row mean $m_q^{(i)}=G^{-1}\sum_jr^{(i,j)}$ measures query action $i$,
and the column mean $m_D^{(j)}=G^{-1}\sum_ir^{(i,j)}$ measures document
bundle $j$. These averages make feedback less dependent on any single
pairing. We then compare each action's mean reward with those of the other
actions in the same role using REINFORCE with a leave-one-out baseline
(REINFORCE-LOO, or \textbf{RLOO}) \citep{kool2019buy,ahmadian2024back}.
For role $c\in\{q,D\}$, the unnormalized advantage is
\begin{equation}
 A_c^{(g)}=m_c^{(g)}-\frac{1}{G-1}\sum_{g'\ne g}m_c^{(g')}
 =\frac{G}{G-1}\bigl(m_c^{(g)}-\bar r\bigr),
 \label{eq:per-side-advantages}
\end{equation}
where $\bar r=G^{-2}\sum_{i,j}r^{(i,j)}$ is the grand mean.
These advantages weight the query and document log-densities in the
surrogate loss:
\begin{equation}
 \mathcal L_{\rm RLOO}=-\frac1G\sum_i A_q^{(i)}
 \log\pi_\theta(e_q^{(i)}\mid q)
 -\frac1G\sum_j A_D^{(j)}
 \log\pi_\theta(E_D^{(j)}\mid D),
 \label{eq:rloo-loss}
\end{equation}
where advantages and sampled actions are held fixed during differentiation.
The bundle log-density sums over valid documents. The ascent estimator is
$\widehat g_{\rm RLOO}=-\nabla_\theta\mathcal L_{\rm RLOO}$.
Conditional on the opposite role's draws, each baseline is independent of
the action it evaluates and has zero expected score-weighted contribution.
Thus $\mathbb E[\widehat g_{\rm RLOO}]=\nabla_\theta\mathcal{J}_\kappa$.
Each rollout batch supplies one on-policy update.

To express this update in encoder parameters, let
$\mathbf{J}_x=\partial h_x/\partial\theta$ be the encoder Jacobian.
For the vMF policy at fixed $\kappa$,
\begin{equation}
 \nabla_\theta\log\pi_\theta(e_x\mid x)=\kappa \mathbf{J}_x^\top e_x,
 \qquad \mathbf{J}_x^\top h_x=0.
 \label{eq:policy-score-main}
\end{equation}
Unit normalization leaves only the action's tangent component contributing
to the gradient; its contribution is weighted by the corresponding
advantage. Appendix~\ref{sec:reler-gradient} gives the gradient derivation
and unbiasedness proof.

\subsection{Variance Reduction by Conditional-Mean Projection}
\label{sec:conditional-projection}

Sampling in a high-dimensional embedding space introduces randomness along
many directions, while a reward based on candidate scores observes only
inner products with the candidates. When the candidate span is small
relative to the embedding dimension, many sampled components are invisible
to the reward but still enter the policy-gradient estimate. They can therefore
add gradient noise without providing useful reward information.
\emph{Conditional-mean projection} (CMP) averages out this residual randomness
while preserving every candidate score. It extends document-space projection
\citep{wang2019variance} to vMF actions for both queries and documents.

\paragraph{Projection as conditional averaging.}
Consider a query action $e$ with fixed candidates. Let $\mathcal{S}$ span those
vectors and the policy mean $h$, and let $P_{\mathcal{S}}$ be its orthogonal projector.
Projection retains every candidate score: $(P_{\mathcal{S}}e)^\top v=e^\top v$.
Let $\mathcal H$ contain all other sampled actions. Conditioning on
$\mathcal H$ and $P_{\mathcal{S}}e$ fixes these scores, leaving only the residual
$(I-P_{\mathcal{S}})e$ random. Because $h\in\mathcal{S}$, the vMF distribution is symmetric
under reflection of this residual, whose conditional mean is therefore zero:
\begin{equation}
 \mathbb E[e\mid\mathcal H,P_{\mathcal{S}}e]=P_{\mathcal{S}}e.
 \label{eq:cp-conditional-mean-main}
\end{equation}
We call this \emph{conditional-mean projection} because the projection equals
the action's conditional mean given the other actions and its retained
projection. When the LOO weight is determined by this same information,
the weighted gradient term is also conditionally averaged. CMP uses $P_{\mathcal{S}}e$
only as a gradient coefficient, without renormalization or changes to
sampling and reward evaluation.

\paragraph{Projection for product rollouts.}
Each reward cell $(i,j)$ uses document bundle $j$ and query draw $i$.
Including the corresponding policy means gives the subspaces
\begin{equation}
 \mathcal{S}_j^q=\operatorname{span}(h_q,e_{d_1}^{(j)},\ldots,e_{d_n}^{(j)}),
 \qquad \mathcal{S}_{im}^d=\operatorname{span}(h_{d_m},e_q^{(i)}).
 \label{eq:cp-spaces}
\end{equation}
The query space retains all candidate directions; each document space
retains its score against query draw $i$. Because these spaces vary across
cells, we keep cellwise LOO weights until projection:
\begin{equation}
 a^q_{ij}=r^{(i,j)}-\frac{\sum_{i'\ne i}r^{(i',j)}}{G-1},\qquad
 a^D_{ij}=r^{(i,j)}-\frac{\sum_{j'\ne j}r^{(i,j')}}{G-1}.
 \label{eq:cell-loo}
\end{equation}
Each baseline holds the opposite role's draw fixed, and all documents in a
bundle share $a^D_{ij}$. Averaging recovers the marginal advantages:
$A_q^{(i)}=G^{-1}\sum_j a^q_{ij}$ and
$A_D^{(j)}=G^{-1}\sum_i a^D_{ij}$.
For reward $R$, the projected estimator, \textbf{RLOO+CMP}, is
\begin{equation}
 \widehat g_{\rm CMP}=\frac{\kappa}{G^2}\sum_{i,j=1}^{G}
 \left[a^q_{ij}\mathbf{J}_q^\top P_{\mathcal{S}_j^q}e_q^{(i)}
 +a^D_{ij}\sum_{m=1}^{n}\mathbf{J}_{d_m}^\top P_{\mathcal{S}_{im}^d}e_{d_m}^{(j)}\right].
 \label{eq:cp-gradient}
\end{equation}
Replacing the projectors by the identity recovers $\widehat g_{\rm RLOO}$.

For bounded rewards determined by the retained candidate scores, CMP
preserves the expected gradient and does not increase any individual
summand's variance along any parameter direction
(Appendices~\ref{sec:cp-details} and~\ref{sec:cp-implementation}).

\subsection{Query Adaptation with a Fixed Index}
\label{sec:fixed-corpus-mode}

In practice, dense retrieval is typically served through an approximate
nearest neighbor search (ANNS) index built over precomputed document embeddings,
which is costly to rebuild whenever the document encoder changes.
In this setting, RELER instead adapts only the query encoder.
We freeze the pretrained document encoder
and keep its precomputed corpus embeddings unchanged. A separate query encoder,
initialized from the same pretrained weights, learns to retrieve against these
fixed vectors without document re-encoding or index rebuilding.

For each training query, we draw a group of $G$ query vectors from its vMF
policy. Each vector retrieves a top-$k$ list from the full index and receives
a reward computed from that list. Query-only RLOO uses the mean of the other
$G-1$ rewards as each action's baseline and updates the query encoder through
the first term of Eq.~\eqref{eq:rloo-loss}. No document sampling or product
rollouts is needed. We do not apply CMP in this
fixed-index setting. Projecting onto the span of the policy mean and
retrieved document vectors preserves the scores of those documents, but
need not preserve the full-index top-$k$ list: documents outside the
retrieved set can cross the retrieval cutoff.

We also anchor the query encoder to its pretrained initialization: for each
training query, we penalize the cosine distance between its current
deterministic embedding and its initial embedding. This limits drift relative
to the frozen index; the loss and its fixed-concentration vMF interpretation
are given in Appendix~\ref{sec:fixed-index-anchor}.

The reward can use retrieval quality, downstream performance, or both.
For RAG, a frozen generator produces an answer from the original query and
each retrieved list. Passage relevance labels and reference answers provide
the two sources of feedback (Section~\ref{sec:reward-design}). The same RLOO update then
trains the query encoder without differentiating through retrieval or
generation. At deployment, the deterministic query embedding $h_q$ replaces
sampled actions and searches the unchanged index.

\subsection{Training Rewards}
\label{sec:reward-design}

We instantiate rewards from relevance labels for retrieval and reference
answers for RAG.

\paragraph{Retrieval rewards.}
From supervision $y$, we compute $R_{\mathrm{nDCG}}$ as nDCG@10 with
exponential gains on $D$ and form positive--negative pairs $\mathcal P_q$
from its labels. With scores $s_p,s_n$, we use
\begin{equation}
 R=R_{\mathrm{nDCG}}+\lambda R_{\mathrm{pair}},\qquad
 R_{\mathrm{pair}}=\frac{1}{|\mathcal P_q|}
 \sum_{(p,n)\in\mathcal P_q}
 \left(\mathbf 1[s_p>s_n]+\tfrac12\mathbf 1[s_p=s_n]\right),
 \quad\lambda=0.5.
 \label{eq:combined-reward}
\end{equation}
The nDCG reward emphasizes graded
ordering near the cutoff; the pairwise term covers the full candidate set.
Appendix~\ref{sec:reward-design-principles} specifies the gains and labels.

\paragraph{Componentwise projection.}
We apply Eq.~\eqref{eq:cp-gradient} to nDCG to obtain
$\widehat g_{\rm CMP}^{\rm nDCG}$. Each pair uses its own LOO weights and
a query subspace spanning the policy mean and document-vector difference,
since its reward depends only on the corresponding score difference.
Averaging the pair estimators gives $\widehat g_{\rm pair}$
(Appendix~\ref{sec:pairwise-estimator}), yielding
\begin{equation}
 \widehat g_{\rm RELER}
 =\widehat g_{\rm CMP}^{\rm nDCG}+\lambda\widehat g_{\rm pair}.
 \label{eq:combined-gradient}
\end{equation}
This is unbiased for the combined reward by linearity; both terms share
the actions, candidates, and backward pass.

\paragraph{Answer feedback for RAG.}
For the fixed-index RAG setting (Section~\ref{sec:fixed-corpus-mode}), let
$L_{10}$ be the retrieved top-ten list and $\hat a$ the frozen generator's
answer from that list. We compare graded retrieval nDCG@10, answer token F1,
and their equal mixture:
\begin{equation}
 R_{\mathrm{RAG}}^{\mathrm{mix}}
 =\tfrac12\,\mathrm{nDCG@10}(L_{10};y_{\mathrm{rel}})
 +\tfrac12\,\mathrm{F1}(\hat a,y_{\mathrm{ans}}),
 \label{eq:rag-mixed-reward}
\end{equation}
where $y_{\mathrm{rel}}$ contains passage relevance grades and
$y_{\mathrm{ans}}$ contains reference answers. The F1 term uses the task's
answer-normalization rules; each single-reward control uses one of the two
terms alone.

\section{Experiments}
\label{sec:experiments}

We evaluate BRIGHT post-training and fixed-index QA adaptation,
with ablations of rewards, rollouts, and gradient estimation.

\subsection{Reasoning-Intensive Retrieval}
\label{sec:main-results}

\subsubsection{Experimental Setup}
\label{sec:experimental-setup}

\paragraph{Baselines.}
We compare RELER with each pretrained encoder and two supervised
objectives: LambdaLoss \citep{wang2018lambdaloss}, which weights pairwise
logistic losses by nDCG@10 swap changes using teacher grades, and InfoNCE
\citep{oord2018representation}, which averages positive-versus-negative
contrastive losses over known positives at temperature 0.03.

\paragraph{Training setup.}
We adapt Qwen3-Embedding-0.6B/4B (Q3E-0.6B/4B) \citep{zhang2025qwen3embedding} and
BGE-M3 \citep{chen2024bgem3} on 4,963 queries from ReasonRank
\citep{liu2025reasonrank}, each with up to 20 candidates. LLM-provided candidate
rankings supply the graded relevance labels used by LambdaLoss and RELER.
All methods train for 113 steps (approximately 3 epochs),
using paired seeds 42, 3407, and 2026.
Q3E-0.6B and BGE-M3 are fully fine-tuned at a learning rate of $5\times10^{-6}$.
For Q3E-4B, all methods use LoRA with a learning rate of $2\times10^{-4}$.
The RELER recipe combines product rollouts
and RLOO+CMP with $G=64$, $\rho=0.70$, and a graded nDCG@10 reward augmented
by pairwise feedback ($\lambda=0.5$). We run all experiments on eight NVIDIA
H100 80GB GPUs. Training time is similar to that of contrastive learning;
embedding sampling and reward computation reuse the encoder outputs.
Full training settings appear in Appendix~\ref{sec:experimental-details}.

\paragraph{Benchmark and evaluation.}
Modern embeddings perform strongly on conventional MTEB/BEIR retrieval but
struggle on reasoning-intensive BRIGHT queries
\citep{zhang2025qwen3embedding,chen2024bgem3,su2025bright}.
We use BRIGHT to test post-training with limited additional supervision.
We report nDCG@10 ($\times$100), averaged equally
over its 12 subsets. Results are means $\pm$ sample SD over three seeds;
pretrained models are evaluated once. Each checkpoint uses both original
queries and the official GPT-4 reasoning queries.

\subsubsection{Main Results}

\begin{table}[!t]
\centering
\caption{BRIGHT nDCG@10 ($\times$100). Subsets show three-seed means;
Avg. is the macro-average $\pm$ SD. Pretrained models are evaluated once.
Bold/underline mark best/second-best scores per model and query setting;
shading marks RELER. TheoQ./TheoT.: TheoremQA questions/theorems.}
\label{tab:main-posttrain}
\begingroup
\compacttablesetup
\setlength{\tabcolsep}{2.6pt}
\begin{tabular}{l*{12}{c}l}
\toprule
 & \multicolumn{7}{c}{\tablehead{StackExchange}} & \multicolumn{2}{c}{\tablehead{Coding}} & \multicolumn{3}{c}{\tablehead{Theorem-based}} & \\
\cmidrule(lr){2-8}\cmidrule(lr){9-10}\cmidrule(lr){11-13}
\tablehead{Method} & \tablehead{Bio.} & \tablehead{Earth.} & \tablehead{Econ.} & \tablehead{Psy.} & \tablehead{Rob.} & \tablehead{Stack.} & \tablehead{Sus.} & \tablehead{Leet.} & \tablehead{Pony} & \tablehead{AoPS} & \tablehead{TheoQ.} & \tablehead{TheoT.} & \tablehead{Avg.} \\
\midrule
\multicolumn{14}{l}{\textit{Original query}} \\
\cmidrule{1-14}
BGE-M3 & 9.5 & 15.3 & 11.9 & 13.2 & 12.1 & 10.6 & 10.2 & \textbf{15.8} & \textbf{15.2} & 1.3 & 8.4 & 4.3 & 10.66 \\
LambdaLoss & 17.2 & 14.1 & 12.4 & 16.2 & 12.8 & 14.0 & 16.6 & \underline{15.0} & 3.5 & \textbf{2.7} & 8.2 & 4.2 & \tablescore{11.40}{0.08} \\
InfoNCE & \underline{18.0} & \underline{17.0} & \underline{14.7} & \underline{20.2} & \underline{15.2} & \underline{14.3} & \underline{16.7} & 10.0 & \underline{8.5} & \underline{2.2} & \underline{10.1} & \underline{4.7} & \tablescore{\underline{12.63}}{0.11} \\
\rowcolor{relerrow}
RELER & \textbf{22.9} & \textbf{25.8} & \textbf{19.4} & \textbf{24.7} & \textbf{17.0} & \textbf{17.0} & \textbf{18.4} & 4.9 & 6.6 & 1.7 & \textbf{10.2} & \textbf{5.4} & \tablescore{\textbf{14.49}}{0.04} \\
\cmidrule{1-14}
Q3E-0.6B & 13.4 & 27.8 & 18.0 & 16.6 & 12.0 & 12.8 & 13.2 & \textbf{14.2} & 0.7 & \textbf{3.3} & 17.2 & \underline{32.1} & 15.10 \\
LambdaLoss & 24.3 & 25.5 & 23.9 & 29.0 & 17.6 & 22.9 & 21.9 & \underline{12.4} & \underline{1.4} & \textbf{3.3} & \textbf{21.0} & 29.0 & \tablescore{19.36}{0.28} \\
InfoNCE & \underline{34.3} & \underline{34.1} & \underline{26.2} & \underline{33.4} & \textbf{21.0} & \underline{28.8} & \textbf{27.4} & 10.2 & 1.2 & \underline{3.1} & \underline{19.7} & \textbf{32.3} & \tablescore{\underline{22.64}}{0.20} \\
\rowcolor{relerrow}
RELER & \textbf{45.9} & \textbf{37.3} & \textbf{28.6} & \textbf{34.0} & \underline{18.4} & \textbf{29.3} & \underline{25.2} & 9.4 & \textbf{2.2} & 2.3 & 19.0 & 28.9 & \tablescore{\textbf{23.38}}{0.17} \\
\cmidrule{1-14}
Q3E-4B & 17.2 & 34.6 & 16.4 & 23.0 & 13.0 & 15.9 & 17.1 & \textbf{21.2} & 2.0 & \underline{3.7} & 19.6 & 40.9 & 18.70 \\
LambdaLoss & 32.3 & 35.5 & 27.5 & 36.0 & 19.9 & 27.0 & 30.8 & \underline{18.5} & 3.6 & \textbf{3.8} & \underline{23.3} & 42.0 & \tablescore{25.01}{0.37} \\
InfoNCE & \underline{48.5} & \underline{47.0} & \underline{32.0} & \underline{42.3} & \textbf{26.7} & \textbf{34.7} & \textbf{35.9} & 15.2 & \underline{6.6} & \textbf{3.8} & \textbf{23.5} & \underline{44.7} & \tablescore{\underline{30.08}}{0.16} \\
\rowcolor{relerrow}
RELER & \textbf{48.9} & \textbf{50.2} & \textbf{32.9} & \textbf{44.3} & \underline{25.8} & \underline{33.4} & \underline{33.6} & 13.8 & \textbf{11.3} & 3.6 & 20.7 & \textbf{46.9} & \tablescore{\textbf{30.46}}{0.55} \\
\midrule
\multicolumn{14}{l}{\textit{GPT-4 reasoning query}} \\
\cmidrule{1-14}
BGE-M3 & 28.3 & 38.1 & 23.1 & \underline{34.8} & 15.0 & \underline{23.9} & 18.4 & \textbf{11.6} & \underline{24.3} & \underline{1.5} & \textbf{14.3} & \textbf{24.0} & \underline{21.45} \\
LambdaLoss & \underline{38.7} & 42.7 & 22.2 & 25.7 & 15.0 & 19.3 & 23.0 & \underline{9.0} & 17.7 & \textbf{2.0} & 8.3 & 6.0 & \tablescore{19.13}{0.10} \\
InfoNCE & 37.1 & \underline{42.9} & \underline{23.5} & 30.8 & \underline{17.4} & 21.8 & \underline{23.8} & 4.3 & 19.8 & 1.1 & \underline{12.0} & 13.0 & \tablescore{20.62}{0.09} \\
\rowcolor{relerrow}
RELER & \textbf{44.7} & \textbf{47.3} & \textbf{25.5} & \textbf{38.2} & \textbf{18.0} & \textbf{24.7} & \textbf{23.9} & 1.5 & \textbf{30.0} & 1.1 & 11.9 & \underline{19.4} & \tablescore{\textbf{23.84}}{0.07} \\
\cmidrule{1-14}
Q3E-0.6B & 27.6 & 36.3 & 24.1 & 28.2 & 13.5 & 17.3 & 19.8 & \textbf{18.6} & 3.2 & 2.1 & 16.9 & 36.8 & 20.38 \\
LambdaLoss & 34.9 & 38.0 & 28.1 & 35.5 & 18.3 & 27.7 & \underline{28.8} & \underline{12.6} & 3.5 & \textbf{3.2} & \textbf{20.8} & 36.3 & \tablescore{23.97}{0.36} \\
InfoNCE & \underline{40.1} & \underline{46.6} & \underline{29.7} & \underline{39.8} & \textbf{21.4} & \textbf{34.9} & \textbf{29.7} & 8.8 & \underline{6.8} & \underline{3.1} & 19.1 & \underline{38.2} & \tablescore{\underline{26.51}}{0.10} \\
\rowcolor{relerrow}
RELER & \textbf{48.1} & \textbf{48.9} & \textbf{30.7} & \textbf{41.1} & \underline{18.8} & \underline{33.0} & 26.5 & 9.8 & \textbf{6.9} & 2.9 & \underline{19.7} & \textbf{39.2} & \tablescore{\textbf{27.15}}{0.04} \\
\cmidrule{1-14}
Q3E-4B & 28.1 & 39.6 & 22.6 & 31.4 & 15.4 & 18.8 & 21.9 & \textbf{21.1} & 2.9 & 3.1 & 20.5 & 38.9 & 22.01 \\
LambdaLoss & 35.4 & 39.3 & 25.5 & 36.3 & 22.5 & 29.8 & 29.8 & \underline{17.8} & 7.9 & \underline{3.3} & \textbf{20.9} & 39.9 & \tablescore{25.69}{0.25} \\
InfoNCE & \underline{51.3} & \underline{49.9} & \underline{31.2} & \underline{42.2} & \underline{25.4} & \textbf{38.8} & \textbf{35.5} & 14.7 & \underline{9.7} & 3.1 & \underline{20.8} & \underline{40.8} & \tablescore{\underline{30.29}}{0.35} \\
\rowcolor{relerrow}
RELER & \textbf{52.5} & \textbf{51.2} & \textbf{32.2} & \textbf{45.3} & \textbf{26.8} & \underline{38.2} & \underline{33.3} & 12.7 & \textbf{14.7} & \textbf{3.9} & 19.4 & \textbf{42.6} & \tablescore{\textbf{31.07}}{0.39} \\
\bottomrule
\end{tabular}

\endgroup
\end{table}

\paragraph{RELER consistently outperforms the baselines across encoders.}
RELER achieves the highest average nDCG@10 on all three backbones with both
original and GPT-4 reasoning queries (Table~\ref{tab:main-posttrain}), ahead of
InfoNCE, LambdaLoss, and the pretrained encoders. Its consistent advantage
over LambdaLoss, which uses the same teacher grades, suggests that graded
supervision alone does not explain the gains. The lead persists with GPT-4
queries, showing that embedding post-training remains beneficial when
query-side reasoning is already provided.

\paragraph{RELER's margins vary across backbones and baselines.}
Relative to InfoNCE, RELER has its largest average gains on BGE-M3:
1.86 points with original queries and 3.22 with GPT-4 reasoning.
For GPT-4 queries, however, InfoNCE falls below the pretrained BGE-M3
encoder; RELER's improvement over that encoder is 2.39 points.
The Qwen backbones gain more over their pretrained encoders but have
smaller margins over InfoNCE. Q3E-4B achieves the highest
macro-average nDCG@10 in both query settings.
The subset scores reveal trade-offs: a higher average does not imply gains
on every domain. This variation may partly reflect how closely each
BRIGHT subset matches the training distribution.

\subsubsection{Ablations and Training Analysis}
\label{sec:ablation-results}
\label{sec:gradient-diagnostics}

\begin{figure}[!htbp]
\centering
\includegraphics[width=\textwidth]{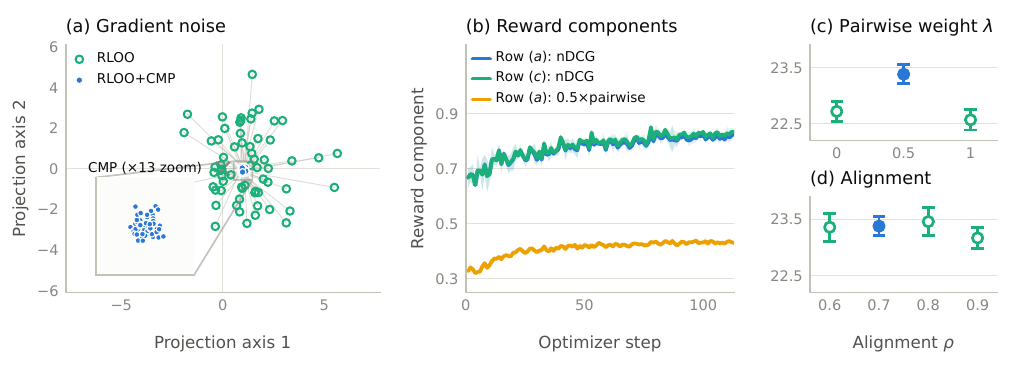}
\caption{\textbf{Training dynamics and sensitivity.}
(a) CMP reduces gradient noise before fine-tuning (inset: zoom).
(b) Pairwise feedback improves BRIGHT despite similar training nDCG reward curves.
(Table~\ref{tab:core-ablations}, rows \textit{(a,c)}).
(c,d) BRIGHT nDCG@10 ($\times$100) is highest at $\lambda=0.5$ and stable across
$\rho=0.60$--$0.90$. Filled points mark defaults; unsmoothed curves and error bars
show three-seed means $\pm$ SD. Sweep details are in Appendix~\ref{sec:pairwise-ablations}.}
\label{fig:gradient-noise}
\label{fig:reward-dynamics}
\label{fig:pairwise-weight}
\label{fig:alignment}
\end{figure}

\paragraph{CMP improves accuracy and reduces seed variation.}
With the combined reward, BRIGHT rises from $21.91\pm0.70$ to $23.38\pm0.17$
(Table~\ref{tab:core-ablations}, \textit{(b)} to \textit{(a)}).
CMP also adds 1.87 points with graded nDCG alone; every paired seed improves
in both comparisons. Matched probes at fixed encoder states show
291--590-fold lower gradient variance (Table~\ref{tab:gradient-probes};
Figure~\ref{fig:gradient-noise}(a)). Since CMP preserves the expected gradient,
these gains are consistent with more effective optimization through reduced
rollout noise. Appendix~\ref{sec:gradient-probe-protocol} details the protocol.

\paragraph{Embedding exploration benefits from product rollouts.}
Product rollouts outperform one-to-one pairing by 2.14 points under graded
nDCG without CMP (\textit{(d)} vs.\ \textit{(e)}). Recombining the same
actions into $G^2$ reward cells lets each query draw receive feedback from
every document bundle, without an additional encoder forward pass.
Query-only and document-only sampling also lower performance
(\textit{(f)} and \textit{(g)}), despite updating the shared encoder.
These results favor exploring both embedding sides and reusing their samples.

\begin{table}[!htbp]
\centering
\caption{BRIGHT ablations (Q3E-0.6B, original queries).
Avg. is nDCG@10 ($\times$100), mean $\pm$ SD over three seeds;
StackEx., Coding, and Theorem average 7, 2, and 3 subsets, respectively.
Shading marks full RELER.
Query-only and document-only sample one embedding side while updating the
shared encoder.
Full results are in Appendix~\ref{sec:additional-ablations}.}
\label{tab:reward-ablation-k0}
\label{tab:core-ablations}
\compacttablesetup
\begin{tabular}{@{}cl*{5}{c}l@{}}
\toprule
\tablehead{} & \tablehead{Reward} & \tablehead{Rollout} & \tablehead{CMP} & \tablehead{StackEx.} & \tablehead{Coding} & \tablehead{Theorem} & \tablehead{Avg.} \\
\midrule
\rowcolor{relerrow}
\textit{(a)} & Graded + pairwise & Product & Yes & 31.25 & 5.79 & 16.73 & \tablescore{\textbf{23.38}}{0.17} \\
\textit{(b)} & Graded + pairwise & Product & No & 28.09 & 5.73 & 18.27 & \tablescore{21.91}{0.70} \\
\midrule
\textit{(c)} & Graded nDCG & Product & Yes & 29.80 & 6.21 & 17.24 & \tablescore{22.72}{0.18} \\
\textit{(d)} & Graded nDCG & Product & No & 26.22 & 5.79 & 18.36 & \tablescore{20.85}{0.68} \\
\textit{(e)} & Graded nDCG & Paired & No & 22.55 & 6.68 & 17.78 & \tablescore{18.71}{0.53} \\
\textit{(f)} & Graded nDCG & Query only & No & 17.92 & 8.64 & 15.85 & \tablescore{15.85}{0.26} \\
\textit{(g)} & Graded nDCG & Document only & No & 22.10 & 5.98 & 18.25 & \tablescore{18.45}{1.56} \\
\midrule
\textit{(h)} & Binary nDCG & Product & Yes & 21.74 & 4.49 & 16.07 & \tablescore{17.45}{0.65} \\
\textit{(i)} & Binary MRR & Product & Yes & 23.42 & 4.69 & 17.83 & \tablescore{18.90}{0.43} \\
\bottomrule
\end{tabular}

\end{table}
\FloatBarrier

\paragraph{Graded rewards outperform binary rewards.}
With product rollouts and CMP fixed, replacing binary labels with teacher grades raises
the nDCG-based result from 17.45 to 22.72 (\textit{(h)} vs.\ \textit{(c)}).
Binary MRR fares better than binary nDCG but remains well below graded
nDCG. Distinguishing levels of candidate relevance matters more here than
the choice between the two binary metrics.

\paragraph{Pairwise feedback brings further gains.}
Adding the pairwise term improves BRIGHT by a further 0.66 points, with
gains at all three seeds (\textit{(c)} to \textit{(a)}).
The two recipes follow similar training nDCG curves, while the pairwise
reward continues to rise (Figure~\ref{fig:reward-dynamics}(b)). The extra
feedback improves the final retrieval score without a clear separation
in training nDCG.

\paragraph{A larger pairwise weight hurts performance.}
Setting $\lambda=1$ reduces BRIGHT to 22.57, slightly below graded-only training
(Figure~\ref{fig:pairwise-weight}(c)). The best tested weight is 0.5.
Because the pairwise reward weights comparisons throughout the candidate
list, a larger coefficient may put too much emphasis on ordering away
from the top ranks favored by nDCG@10.

\paragraph{Performance is stable across alignment values.}
BRIGHT varies little over $\rho=0.60$--$0.90$
(Figure~\ref{fig:alignment}(d)). The best mean occurs at 0.80, but its
advantage over the default 0.70 changes sign across seeds. Within this
range, the method is not sensitive to a precise alignment value.
Lower alignment allows samples to move farther from the encoder output;
higher alignment keeps them closer. Neither direction yields a consistent
improvement over the default.
The full sweep spans only 0.29 points, much less than the effects of CMP
or rollout design under the same training budget.

\FloatBarrier

\begin{table}[!htbp]
\centering
\caption{Fixed-index QA ($\times$100), seed 42.
In-domain/out-of-domain average 2/5 datasets; Avg. covers all seven.
Bold marks column bests. Per-dataset results are in
Appendix~\ref{sec:rag-details}.}
\label{tab:rag-main}
\begingroup
\tablesetup
\setlength{\tabcolsep}{3.0pt}
\begin{tabular}{@{}l*{9}{c}@{}}
\toprule
 & \multicolumn{3}{c}{\tablehead{In-domain}} & \multicolumn{3}{c}{\tablehead{Out-of-domain}} & \multicolumn{3}{c}{\tablehead{Avg.}} \\
\cmidrule(lr){2-4}\cmidrule(lr){5-7}\cmidrule(l){8-10}
\tablehead{Method} & EM & F1 & Hit@10 & EM & F1 & Hit@10 & EM & F1 & Hit@10 \\
\midrule
Q3E-0.6B & 33.79 & 44.17 & 65.67 & 31.71 & 39.40 & 51.94 & 32.30 & 40.76 & 55.87 \\
InfoNCE & 33.63 & 44.07 & 64.30 & 30.19 & 37.76 & 46.72 & 31.18 & 39.56 & 51.74 \\
RLOO, graded nDCG & 34.66 & 45.08 & \textbf{67.25} & 31.38 & 39.08 & 51.97 & 32.32 & 40.80 & 56.34 \\
RLOO, answer F1 & 34.71 & \textbf{45.16} & 66.77 & 31.38 & 39.14 & \textbf{52.61} & 32.33 & 40.86 & \textbf{56.66} \\
RLOO, nDCG + F1 & \textbf{34.82} & 45.14 & 66.87 & \textbf{32.21} & \textbf{39.48} & 52.19 & \textbf{32.96} & \textbf{41.10} & 56.38 \\
\bottomrule
\end{tabular}

\endgroup
\end{table}

\subsection{Retrieval-Augmented Generation}
\label{sec:rag-results}

\subsubsection{Experimental Setup}
We build a Q3E-0.6B index over FlashRAG's 2018 Wikipedia passage corpus
\citep{jin2025flashrag}. The index and Qwen2.5-7B-Instruct generator remain
frozen, so all adaptation comes from updating the query encoder.
We train on DPR-labeled NQ and FlashRAG HotpotQA for one epoch with seed 42
and query-anchor coefficient 0.5. Human-positive NQ passages and annotated
HotpotQA supporting facts are aligned to the corpus as binary relevance
labels. The five out-of-domain datasets are PopQA, TriviaQA,
2WikiMultihopQA, MuSiQue, and Bamboogle; none supplies training queries.
InfoNCE trains on a fixed offline candidate pool; query-only RLOO searches
the full index with graded nDCG@10, answer F1, or their equal mixture as
reward, without CMP. InfoNCE negatives are mined from the constructed index;
offline mining and online retrieval are part of the respective training
methods being compared. Evaluation retrieves once per question, passing the
top-10 passages to the generator. We report answer EM/F1 and
answer-containing-passage Hit@10, macro-averaged across datasets.
Full details are in Appendix \ref{sec:rag-training-details}.

\subsubsection{Results and Analysis}

\paragraph{RLOO improves fixed-index retrieval and QA.}
All three RLOO recipes improve Hit@10 and answer EM/F1 over InfoNCE
(Table~\ref{tab:rag-main}). With the document index and generator frozen,
these gains show that adapting the query encoder can improve the context
delivered to an existing RAG pipeline. The mixed reward also exceeds the
pretrained encoder on all three seven-dataset averages, although its F1
margin is smaller than the gap over InfoNCE. The reference matters because
InfoNCE falls below the pretrained encoder, while the RLOO recipes recover
or exceed its average scores. RLOO searches the full index, whereas InfoNCE
uses a fixed candidate pool, so their gap reflects both objective and
candidate access.

\paragraph{Answer feedback captures utility beyond retrieval hits.}
Answer-F1 reward slightly improves average EM/F1 over graded nDCG, and
their mixture gives the best overall answer scores. Answer-F1 has the
highest overall Hit@10 but trails the mixture in EM/F1; graded nDCG and
the mixture also have nearly identical Hit@10 despite different answer
scores. Hit@10 counts any answer-containing top-ten passage and ignores passage order,
distracting context, and evidence removed by the generator's input limit.
Reward choice can thus improve the utility of the generator's context even
when answer containment changes little.

\FloatBarrier

\section{Conclusion}

RELER aligns embedding models with retrieval and downstream task rewards through
vMF exploration in embedding space during training, while retaining deterministic
vector retrieval at inference. Conditional-mean projection removes
reward-invisible gradient noise without changing the expected fixed-candidate
gradient. Across BRIGHT backbones, RELER generally improves average retrieval
over InfoNCE, although gains vary by query form and domain. With a fixed
document index and generator, adapting only the query encoder improves
both retrieval performance and end-to-end answer quality.
These results show the potential of embedding-space RL for retrieval post-training,
and motivate future work on adaptive exploration policies and multi-step
retrieval for search agents.

\bibliographystyle{iclr2027_conference}
\bibliography{references}

\appendix
\clearpage
\raggedbottom

Appendix~\ref{sec:theoretical-analysis} gives the derivations and proofs,
and Appendix~\ref{sec:implementation-details} describes the implementation.
Experimental settings and results are grouped by study: BRIGHT in
Appendix~\ref{sec:experimental-details}, fixed-index RAG in
Appendix~\ref{sec:rag-appendix}, and training from a general-purpose LLM in
Appendix~\ref{sec:initialization-study}. Extended related work appears in
Appendix~\ref{sec:extended-related-work}.

\section{Theoretical Analysis}
\label{sec:theoretical-analysis}

We first derive the product-rollout policy gradient, then prove CMP's
unbiasedness and per-term variance guarantee, and finally specialize the
estimator to pairwise rewards. Throughout, the training record, candidate
masks, encoder state, and concentration are fixed for each update.

\subsection{Gradient Estimation for RELER}
\label{sec:reler-gradient}
\label{sec:vmf-policy-gradient}
\label{sec:product-decomposition}

Fix a training record $(q,D,y)$ and concentration $\kappa$. The objective is
$\mathcal{J}_\kappa=\mathbb E[R(e_q,E_D;y)]$ under independent query and
document vMF actions. The labeled candidate set $D$ is fixed;
candidate construction is outside the derivative studied here.
For a sampled action $e_x$, the vMF log-density is
$\log\pi_\theta(e_x\mid x)=\log C_d(\kappa)+\kappa h_x^\top e_x$.
Since the normalizer is independent of $\theta$ at fixed $\kappa$,
\begin{equation}
 \nabla_\theta\log\pi_\theta(e_x\mid x)=\kappa \mathbf{J}_x^\top e_x,
 \qquad \mathbf{J}_x=\frac{\partial h_x}{\partial\theta}.
 \label{eq:vmf-score}
\end{equation}
Sampled actions are held fixed when differentiating the log-density;
gradients pass through the encoder means. The surrogate can omit
$\log C_d(\kappa)$ without changing its gradient.

\paragraph{Unit normalization and tangent gradients.}
The encoder output satisfies $h_x^\top h_x=1$ for every $\theta$.
Differentiating this identity by the chain rule gives
\begin{equation}
 0=\nabla_\theta(h_x^\top h_x)
  =2\left(\frac{\partial h_x}{\partial\theta}\right)^\top h_x
  =2\mathbf{J}_x^\top h_x.
 \label{eq:normalized-jacobian}
\end{equation}
Thus $\mathbf{J}_x^\top h_x=0$: any first-order change of a unit embedding
is orthogonal to its current direction. Decompose the sampled action into
its radial and tangent components,
$e_x=(h_x^\top e_x)h_x+[e_x-(h_x^\top e_x)h_x]$.
The radial term vanishes under $\mathbf{J}_x^\top$, so
\begin{equation}
 \nabla_\theta\log\pi_\theta(e_x\mid x)
 =\kappa\mathbf{J}_x^\top
   \bigl[e_x-(h_x^\top e_x)h_x\bigr].
 \label{eq:tangent-policy-score}
\end{equation}
The bracketed vector is orthogonal to $h_x$ and therefore lies in the
tangent space of the unit sphere at $h_x$. This establishes both identities
in Eq.~\eqref{eq:policy-score-main} and explains why radial action components
contribute no encoder gradient.

\paragraph{Query and document contributions.}
RELER samples the query and every document independently, although their
means share encoder parameters. The document-bundle log-density is therefore
\begin{equation}
 \log\pi_\theta(E_D\mid D)=\sum_{m=1}^{n}
 \log\pi_\theta(e_{d_m}\mid d_m).
 \label{eq:slate-log-prob}
\end{equation}
The sum covers valid documents without list-length averaging. For
$X\sim\pi_\theta(\cdot\mid q)$ and $Y\sim\pi_\theta(\cdot\mid D)$,
the two contributions to $\nabla_\theta \mathcal{J}_\kappa$ are
\begin{equation}
 \begin{aligned}
 g_q&=\mathbb E_X\!\left[
   \mathbb E_Y[R(X,Y)]\nabla_\theta\log\pi_\theta(X\mid q)\right],\\
 g_D&=\mathbb E_Y\!\left[
   \mathbb E_X[R(X,Y)]\nabla_\theta\log\pi_\theta(Y\mid D)\right],
 \end{aligned}
 \qquad \nabla_\theta \mathcal{J}_\kappa=g_q+g_D.
 \label{eq:product-policy-gradient}
\end{equation}
The reward table's row and column means estimate these inner expectations:
each query action is evaluated across document bundles, and each bundle
across query actions. Pairing $G$ draws from each role reuses $2G$ component
actions for $G^2$ correlated reward evaluations
\citep{kuntz2022product}. Shared encoder parameters do not make the draws
dependent, and the reward evaluations are not independent samples.

\paragraph{Leave-one-out feedback.}
For a query action $i$, condition on all document-bundle draws. Its baseline
uses only the other query draws and is independent of action $i$; the action's
log-density gradient has conditional mean zero. The baseline therefore has
zero expected score-weighted contribution. The same argument applies to a
document bundle $j$ after conditioning on the query draws and excluding
bundle $j$ from its baseline. Applying this argument to the surrogate in
Eq.~\eqref{eq:rloo-loss} gives
\begin{equation}
 \mathbb E[\widehat g_{\rm RLOO}]=g_q+g_D=\nabla_\theta \mathcal{J}_\kappa.
 \label{eq:reler-rloo-unbiased}
\end{equation}
This is the standard RLOO baseline argument applied to RELER's paired actions
\citep{williams1992simple,kool2019buy,ahmadian2024back}. The statement is
conditional on the fixed record, candidate masks, and concentration.
RELER uses no sample-dependent reward standardization or document-density
averaging; either modification can change the expected gradient. Each
rollout batch supplies one on-policy encoder update.

\subsection{Conditional-Mean Projection: Unbiasedness and Variance}
\label{sec:cp-details}

CMP replaces each sampled action in a weighted gradient term by its
projection, retaining the same cellwise LOO weight. For query draw $i$,
document bundle $j$, and document $m$, the projection spaces are
\begin{equation}
 \mathcal{S}_j^q=\operatorname{span}(h_q,e_{d_1}^{(j)},\ldots,e_{d_n}^{(j)}),
 \qquad \mathcal{S}_{im}^d=\operatorname{span}(h_{d_m},e_q^{(i)}),
 \label{eq:cp-spaces-appendix}
\end{equation}
These spaces contain the policy mean and every direction needed to retain
the current action's candidate scores.

\begin{proposition}[Unbiasedness and variance reduction]
\label{prop:conditional-projection}
Fix the encoder state, record, candidate masks, and
concentration $\kappa$. Assume independent vMF actions and a bounded reward depending
on each action only through the candidate inner products retained by
Eq.~\eqref{eq:cp-spaces-appendix}. The CMP estimator is unbiased:
$\mathbb E[\widehat g_{\rm CMP}]=\mathbb E[\widehat g_{\rm RLOO}]
=\nabla_\theta \mathcal{J}_\kappa$.
Each projected query summand $(i,j)$ and document summand $(i,j,m)$ has
no greater variance along any parameter direction than its unprojected
counterpart.
\end{proposition}

\paragraph{Proof of Proposition~\ref{prop:conditional-projection}.}
At a fixed encoder state, the means and Jacobians are deterministic.
Consider any one query or individual-document summand of the estimator.
Let $e$ be its current action, $\mathcal{S}$ its subspace,
$\mathbf{J}$ its Jacobian, and $a$ its cellwise LOO weight. Let $\mathcal H$ contain
all other sampled actions and define
$\mathcal G=\sigma(\mathcal H,P_{\mathcal{S}}e)$.
The subspace is $\mathcal H$-measurable and does not depend on $e$.
By independence, conditional on $\mathcal H$, $e$ still follows its vMF law.

Write $Q=2P_{\mathcal{S}}-I$. This orthogonal map fixes $\mathcal{S}$ and negates $\mathcal{S}^\perp$.
Because $h\in\mathcal{S}$, $h^\top Qe=h^\top e$, and $Q$ preserves both surface
measure and the vMF density. It also leaves $P_{\mathcal{S}}e$ unchanged. Hence the
conditional law of $(I-P_{\mathcal{S}})e$ given $\mathcal G$ is symmetric about zero,
which proves
\begin{equation}
 \mathbb E[e\mid\mathcal G]=P_{\mathcal{S}}e.
 \label{eq:cp-conditional-mean}
\end{equation}
For a query term, every score in bundle $j$, including every valid frozen
candidate, is determined by $P_{\mathcal{S}_j^q}e_q^{(i)}$. The baseline excludes
query action $i$, so it is already determined by $\mathcal H$.
For a document term, condition on all other documents in the current bundle
as well as all other actions. The current document affects the reward only
through $e_q^{(i)\top}e_{d_m}^{(j)}$; the baseline excludes bundle $j$.
Thus $a$ is $\mathcal G$-measurable in both cases.

For $Z=\kappa a\mathbf{J}^\top e$, Eq.~\eqref{eq:cp-conditional-mean} yields
$Z_{\rm CMP}=\mathbb E[Z\mid\mathcal G]=\kappa a\mathbf{J}^\top P_{\mathcal{S}}e$.
This is the Rao--Blackwell construction \citep{casella1996rao} applied to
the entire weighted gradient term.
Taking expectations and summing the $G^{-2}$-weighted terms proves
$\mathbb E[\widehat g_{\rm CMP}]=\mathbb E[\widehat g_{\rm RLOO}]$.
The preceding LOO derivation identifies their common expectation with
$\nabla_\theta \mathcal{J}_\kappa$.
For bounded rewards, the law of total covariance also gives
\begin{equation}
 \operatorname{Cov}(Z)-\operatorname{Cov}(Z_{\rm CMP})
 =\mathbb E[\operatorname{Cov}(Z\mid\mathcal G)]\succeq0.
 \label{eq:cp-term-covariance}
\end{equation}
The conditioning $\mathcal G$ differs between summands. Consequently this
identity does not order the covariance of their sums. \hfill$\square$

\paragraph{Scope of the guarantee.}
The result concerns the stated estimator with candidates and concentration
fixed within each rollout table. All reward-relevant frozen directions must
be retained. In particular, a subspace containing only returned full-corpus
top-$k$ hits need not preserve the retrieved list and does not satisfy this
condition. The variance guarantee is per term; correlations between terms
prevent an automatic ordering of the summed gradient covariance.

\subsection{Pairwise Conditional-Mean Projection}
\label{sec:pairwise-estimator}

For pair $(p,n)\in\mathcal P_q$ and document bundle $j$, let
$\delta_{pn}^{(j)}=v_p^{(j)}-v_n^{(j)}$. Each $v$ is a sampled own-query
vector $e_d^{(j)}$ or a fixed vector $\rho h_d$, where
$\rho=\mathbb E[h_d^\top e_d]$. Define
$r_{ij}^{pn}=\mathbf 1[e_q^{(i)\top}\delta_{pn}^{(j)}>0]
+\frac12\mathbf 1[e_q^{(i)\top}\delta_{pn}^{(j)}=0]$.
Its LOO weights are
\begin{equation}
 a^{q,pn}_{ij}=r_{ij}^{pn}-\frac{1}{G-1}\sum_{i'\ne i}r_{i'j}^{pn},
 \qquad
 a^{D,pn}_{ij}=r_{ij}^{pn}-\frac{1}{G-1}\sum_{j'\ne j}r_{ij'}^{pn}.
\end{equation}
Let $\mathcal I_{pn}$ denote the pair's sampled own-query endpoints. With
$\mathcal{S}_{j,pn}^{q}=\operatorname{span}(h_q,\delta_{pn}^{(j)})$ and document spaces
$\mathcal{S}_{im}^{d}=\operatorname{span}(h_{d_m},e_q^{(i)})$, the estimator is
\begin{equation}
 \widehat g_{\rm pair}=\frac{\kappa}{G^2|\mathcal P_q|}
 \sum_{(p,n)\in\mathcal P_q}\sum_{i,j}
 \left[a^{q,pn}_{ij}\mathbf{J}_q^\top P_{\mathcal{S}_{j,pn}^{q}}e_q^{(i)}
 +a^{D,pn}_{ij}\sum_{m\in\mathcal I_{pn}}
 \mathbf{J}_{d_m}^\top P_{\mathcal{S}_{im}^{d}}e_{d_m}^{(j)}\right].
 \label{eq:pairwise-cp-gradient}
\end{equation}
For an empty pair set this is zero. Each query projection retains its pair's
score difference; each document projection retains that endpoint's score.
The symmetry argument of Proposition~\ref{prop:conditional-projection}
therefore applies to every term. Actions of documents outside the pair have
zero expected score-weighted contribution and can be omitted. It follows that
$\mathbb E[\widehat g_{\rm pair}]=\nabla_\theta\mathbb E[R_{\rm pair}]$,
conditional on the fixed record, detached additions, and $\kappa$.
For $R=R_{\rm nDCG}+\lambda R_{\rm pair}$, combining the listwise and
pairwise estimators gives
$\mathbb E[\widehat g_{\rm RELER}]=\nabla_\theta \mathcal{J}_\kappa$ for the combined
reward. Linearity permits shared actions
across reward components; the termwise variance guarantee does not order
the covariance of their sum.

\FloatBarrier
\section{Implementation Details}
\label{sec:implementation-details}

We describe embedding sampling, reward evaluation, and gradient projection,
then combine them into the complete training algorithm.

\subsection{Exploration and Embedding Sampling}
\label{sec:sampling-settings}
\label{sec:exploration-scale}

RELER specifies exploration through the expected alignment of a sampled
embedding with its encoder output:
\begin{equation}
 \rho=A_d(\kappa)=\mathbb E[h_x^\top e_x]
 =\frac{I_{d/2}(\kappa)}{I_{d/2-1}(\kappa)},
\end{equation}
where $I_\nu$ is the modified Bessel function of the first kind. We solve
for $\kappa$ at the embedding dimension and use it in the sampler below.

\paragraph{Exact vMF sampling.}

We sample vMF actions using the rejection sampler of
\citet{wood1994simulation}, shown in Algorithm~\ref{alg:vmf-sampling}.
Each action decomposes as $e=wh+\sqrt{1-w^2}\,v$, where $w=h^\top e$
sets its alignment with the mean and $v$ is a uniformly sampled unit
direction orthogonal to $h$. Rejection sampling generates $w$; projecting
and normalizing an independent Gaussian vector generates $v$.

\begin{algorithm}[htbp]
\caption{Sampling a group of vMF embedding actions}
\small
\label{alg:vmf-sampling}
\begin{algorithmic}[1]
\Require Unit mean $h\in\mathbb S^{d-1}$, $d\geq3$; concentration $\kappa>0$; group size $G$
\Ensure Independent actions $e^{(1)},\ldots,e^{(G)}\sim\mathrm{vMF}(h,\kappa)$
\State $a\gets(d-1)/2$
\State $b\gets\dfrac{d-1}{2\kappa+\sqrt{4\kappa^2+(d-1)^2}}$
\State $w_0\gets(1-b)/(1+b)$; $c\gets\kappa w_0+(d-1)\log(1-w_0^2)$
\For{$g=1,\ldots,G$}
  \Repeat \Comment{Sample the alignment coefficient}
    \State $z\sim\operatorname{Beta}(a,a)$; $u\sim\operatorname{Uniform}(0,1)$
    \State $w\gets\dfrac{1-(1+b)z}{1-(1-b)z}$
  \Until{$\kappa w+(d-1)\log(1-w_0w)-c\geq\log u$}
  \State $\xi\sim\mathcal N(0,I_d)$ \Comment{Sample an independent tangent direction}
  \State $v\gets\xi-(h^\top\xi)h$; $v\gets v/\|v\|$
  \State $e^{(g)}\gets wh+\sqrt{1-w^2}\,v$
\EndFor
\State \Return $e^{(1)},\ldots,e^{(G)}$
\end{algorithmic}
\end{algorithm}

We apply this procedure independently to each query and document mean.
The implementation batches all pending draws, evaluates the rejection test
in FP64, and constructs tangent directions and actions in FP32 before
casting to the input dtype. It raises an error if any draw remains
unaccepted after 256 rejection rounds. Sampled actions are held fixed in
the policy-gradient update.

In contrast,
$\operatorname{normalize}(h+\sigma\epsilon)$ follows a projected-normal law.
Scoring such samples with a vMF density does not give the score-function
gradient of their sampling law. We include
that approximation only as a sampling/scoring mismatch diagnostic, using the
nominal local scale $\sigma=1/\sqrt{\kappa}$.

\subsection{Reward Definitions}
\label{sec:reward-design-principles}

In our implementation, $y$ collects teacher-assigned relevance grades and
the original binary relevance annotations. Let $c_1,\ldots,c_M$ be the
candidates for a query, with grades $\ell_m$ and scores $s_m$.
Sorting by score gives a permutation $\pi$.
For cutoff $k$ and $k'=\min(k,M)$,
\begin{equation}
  \mathrm{nDCG@}k=
  \frac{\sum_{t=1}^{k'}(2^{\ell_{\pi_t}}-1)/\log_2(t+1)}
       {\max_{\pi'}\sum_{t=1}^{k'}(2^{\ell_{\pi'_t}}-1)/\log_2(t+1)}.
  \label{eq:ndcg-reward}
\end{equation}
Zero ideal DCG gives zero reward. The pool contains only valid own-query
candidates. Teacher ranks map to grades $\ell_m$: 3 for
rank 1, 2 for ranks 2--5, 1 for ranks 6--10, and 0 otherwise.
The main reward is $R_{\rm nDCG}+\lambda R_{\rm pair}$ with $\lambda=0.5$.
The pairwise term averages a score of 1 for a positive ranked above a
negative, $1/2$ for a tie, and 0 otherwise. Positive membership comes from
the original binary annotations in $y$, not a threshold on teacher grades.
A query with no valid pair receives zero pairwise reward and remains in
the batch average.
For binary nDCG and MRR controls, labels indicate all annotated positives.
MRR@10 is the reciprocal rank of the first positive within the top ten,
or zero if none occurs. All rewards are evaluated exactly; no differentiable
rank relaxation is used.

\paragraph{Exploration and discrete feedback.}
\label{sec:reward-resolution}
For the retrieval rewards used by RELER, changing embeddings provides feedback
only when it changes a reward-relevant ordering. If every pairing in the
reward table has the same value, all marginal and cellwise LOO weights vanish.
Both RLOO and CMP then give zero updates for that reward component: projection
can remove gradient noise but cannot create a missing reward difference.
Concentrating actions too tightly around their means can make such constant
reward tables more frequent; broader exploration changes candidate orderings
more often but can move retrieval behavior farther from deterministic deployment.

RELER uses graded nDCG together with binary pairwise feedback. Graded labels
make more swaps reward-relevant, while an individual positive--negative pair
can change order even when nDCG remains constant. The pairwise term can
therefore supply feedback in those cases.

\subsection{Conditional-Mean Projection Implementation}
\label{sec:cp-implementation}
Projection bases use the current encoder means and sampled actions.
We detach the bases and coefficients, then apply the resulting gradient
through the live unit means.

For listwise rewards, the query subspace spans the policy mean, sampled
documents in the current bundle. The
implementation projects the query weighted sum separately for each bundle.
With at most 20 own-query candidates and the policy mean, the listwise
query subspace has dimension at most 21, compared with $d=1024$.
We construct the basis with reduced SVD in FP32, setting padded and masked
candidate columns to zero. For a $d\times w$ basis-input matrix with largest
singular value $\sigma_{\max}$, we retain singular vectors whose singular
values exceed $\max(d,w)\epsilon_{\rm FP32}\sigma_{\max}$. This handles
duplicate or linearly dependent columns without adding arbitrary directions.
For documents, write
$u_{im}=e_q^{(i)}-(h_{d_m}^\top e_q^{(i)})h_{d_m}$. The tangent projection is
$u_{im}u_{im}^\top e_{d_m}^{(j)}/\|u_{im}\|^2$; the radial component vanishes
under the normalized mean's Jacobian. We form this tangent direction in FP64
to avoid cancellation for nearly parallel vectors, normalize it when its norm
exceeds $10^{-12}$, and set it to zero otherwise; the subsequent contractions
use FP32. The implemented coefficient retains the radial component, which
normalization removes during backpropagation. CMP requires no additional
encoder forwards or reward evaluations.

For each pairwise reward, the query subspace spans the policy mean and
the positive--negative document-vector difference, giving dimension at
most two. We project each
pair before averaging, rather than applying one listwise projector to the
aggregate pair reward. The implementation processes pairs in small blocks,
reuses the product scores, and accumulates query and document coefficients
before one encoder backward pass.
Linearity allows us to sum the query coefficients over query draws before
projecting within each document bundle, and to sum document coefficients
over document draws before projecting for each query draw. These contractions
implement Eq.~\eqref{eq:cp-gradient} without materializing a
$B\times G\times G\times n\times d$ tensor, where $B$ is the microbatch size.
Actions, LOO weights, and projectors are detached throughout; gradients
flow only through the live normalized encoder means. Projected coefficients
are not renormalized to unit length.

\subsection{Query--Document Post-Training Algorithm}
\label{sec:training-algorithm}

Algorithm~\ref{alg:query-slate} assembles the main recipe into one
training loop using each query's labeled documents. Each $\widehat g^{(q)}$ estimates the full shared-encoder
gradient for record $q$. In the pseudocode, $\operatorname{sg}$ denotes
stop-gradient, and \textsc{SampleVMF} is Algorithm~\ref{alg:vmf-sampling}.

\textsc{LOO} subtracts other-row means for query weights and other-column
means for document weights. \textsc{ListCMP} and \textsc{PairCMP} form the
listwise and individual pair gradients using the current actions and means.
The combined batch gradient requires one backward pass; AdamW receives its
negative for ascent. An empty pair set leaves only the listwise contribution.

\begin{algorithm}[H]
\small
\caption{RELER: embedding post-training with retrieval feedback}
\label{alg:query-slate}
\begin{algorithmic}[1]
\Require Encoder $f_{\theta_0}$; minibatches of records $(q,D,y)$
\Require $G=64$, $\rho=0.70$, $\lambda=0.5$
\Ensure Deterministic retriever $f_\theta$
\State $\theta\gets\theta_0$; $\kappa\gets A_d^{-1}(\rho)$
\For{each minibatch $\mathcal B$}
  \State $\{h_x\}\gets f_\theta(\mathcal B)$ \Comment{Encode queries and documents once}
  \For{each $(q,D,y)\in\mathcal B$}
    \State $e_q^{(1:G)}\gets\operatorname{SampleVMF}(h_q,\kappa,G)$
    \State $e_d^{(1:G)}\gets\operatorname{SampleVMF}(h_d,\kappa,G)$ for each $d\in D$
    \State $S=(s_m^{(i,j)})\gets\operatorname{Scores}(e_q^{(1:G)},E_D^{(1:G)})$
    \State $\mathbf R\gets[\operatorname{nDCG@10}(S_{ij,:},y)]_{i,j=1}^{G}$
    \State $(\mathbf a^q,\mathbf a^D)\gets\operatorname{LOO}(\mathbf R)$ \Comment{Cellwise leave-one-out weights}
    \State $\widehat g^{(q)}\gets\operatorname{ListCMP}(\mathbf a^q,\mathbf a^D)$ \Comment{Listwise gradient}
    \State $\mathcal P_q\gets\operatorname{Pairs}(y,D)$
    \For{each $(p,n)\in\mathcal P_q$}
      \State $\mathbf R^{pn}\gets[\mathbf 1[s_p^{(i,j)}>s_n^{(i,j)}]
             +\tfrac12\mathbf 1[s_p^{(i,j)}=s_n^{(i,j)}]]_{i,j=1}^{G}$
      \State $(\mathbf a^{q,pn},\mathbf a^{D,pn})\gets\operatorname{LOO}(\mathbf R^{pn})$
      \State $\widehat g^{(q)}\gets\widehat g^{(q)}+\dfrac{\lambda}{|\mathcal P_q|}
             \operatorname{PairCMP}(\mathbf a^{q,pn},\mathbf a^{D,pn};p,n)$
    \EndFor
  \EndFor
  \State $\theta\gets\operatorname{AdamW}\!\left(\theta,
         -\dfrac{1}{|\mathcal B|}\sum_{q\in\mathcal B}\widehat g^{(q)}\right)$
         \Comment{One optimizer step}
\EndFor
\State \Return $f_\theta$
\end{algorithmic}
\end{algorithm}

\clearpage
\section{BRIGHT Experiments}
\label{sec:experimental-details}
\label{sec:training-evaluation-details}
\label{sec:supplementary-experiments}

This section collects the BRIGHT training and evaluation protocol,
reference baselines, complete ablation results, and gradient measurements.

\subsection{Experimental Setup}

\subsubsection{Data and Preprocessing}
ReasonRank preprocessing removes 133 possible BRIGHT overlaps, 1,546
MS MARCO records, 61 relevance-conflict records, 17 records without negatives,
and one duplicate query from 6,721 records. The remaining 4,963 queries
contain 15,830 positive occurrences. Text deduplication preserves all known
positives; annotations and teacher ranks remain separate supervision sources.
Source-wise batching retains 4,896 queries at a per-device batch size of 16,
reshuffling within sources each epoch. Training seeds pair the data order
across methods; embedding exploration and negative sampling use separate
random-number streams. There is no internal development split. Overlap
screening does not establish complete semantic decontamination.

\subsubsection{Representation and Evaluation Settings}
Training and evaluation use the same model and tokenizer initialization.
For Qwen3-Embedding, we reserve one position when truncating text and append a single terminal
token, then pool its hidden state. Last-token pooling, normalization, and
similarity computation use FP32. Queries contain a task instruction followed
by the query text; documents use raw text. The backbone uses BF16 during
training and FP16 during evaluation. Training query/document limits are
512/1024 tokens; external evaluation permits 8192 tokens. Deterministic
retrieval uses unscaled cosine similarity.

BGE-M3 uses its dense encoder with CLS pooling, right padding, no appended
terminal token, and no query/document prefixes. Training lengths
and the 8192-token evaluation limit follow the same budget as Qwen3.
Its nine training runs use the same prepared ReasonRank data, global batch
128, 113 steps, learning rate $5\times10^{-6}$, and three seeds as the
Q3E-0.6B comparison, with full-encoder fine-tuning. Only dense retrieval is
evaluated; BGE-M3's other retrieval modes are not used.

Both query settings use the same BRIGHT corpus, relevance labels.
\footnote{\href{https://huggingface.co/datasets/xlangai/BRIGHT}{BRIGHT dataset}.}
For GPT-4 reasoning, we use the released query field without human-annotated reasoning.

\subsubsection{Baseline Objectives}
InfoNCE treats each positive separately, excluding other positives from its
denominator, then averages across positives and queries:
\begin{equation}
 \mathcal L_{\rm CL}(q)=-\frac1{|P_q|}\sum_{p\in P_q}
 \log\frac{\exp(s(q,p)/\tau)}
 {\exp(s(q,p)/\tau)+\sum_{n\in N_q}\exp(s(q,n)/\tau)},\qquad\tau=0.03.
 \label{eq:multi-positive-cl}
\end{equation}
InfoNCE uses the full valid cross-device pool, with cross-query document
gradients. LambdaLoss uses teacher grades and normalizes its
weighted logistic sum by the total swap weight within each query. Swap weights
are detached absolute nDCG@10 changes, with logistic scale $1/0.03$ on cosine
scores.

\subsubsection{Training Hyperparameters}
Table~\ref{tab:hyperparams} summarizes the training configurations.

\begin{table}[htbp]
\centering
\caption{Training settings for Q3E-0.6B BRIGHT adaptation.
All 4B BRIGHT methods use LoRA (rank 16) with a learning rate of
$2\times10^{-4}$.}
\label{tab:hyperparams}
\tablesetup
\begin{tabular}{ll}
\toprule
Setting & Q3E-0.6B BRIGHT adaptation \\
\midrule
Trainable parameters & Full shared encoder \\
Optimizer / learning rate & AdamW / $5\times10^{-6}$ \\
Global / per-device batch & 128 / 16 (8 devices) \\
Training steps & 113 \\
Schedule / warm-up fraction & Linear / 0.03 \\
Weight decay & 0.01 \\
Training seeds & 42, 3407, 2026 \\
RL group size / alignment & 64 / 0.70 \\
RL reward & Graded nDCG + pairwise \\
Checkpoint selection & Final step \\
\bottomrule
\end{tabular}
\end{table}

BRIGHT adaptation disables gradient clipping, reference-policy KL, and
auxiliary contrastive losses. Its main reward uses $\lambda=0.5$.
The shared step budget fixes query exposure across methods.

\paragraph{Exploration settings.}
\label{sec:exploration-settings}

The main BRIGHT recipe fixes alignment at $\rho=0.70$ and solves
$A_d(\kappa)=\rho$ for the model's embedding dimension $d$.
The alignment sweep in Figure~\ref{fig:alignment}(d) uses
$\rho\in\{0.60,0.70,0.80,0.90\}$, fixed throughout each run.
The implementation logs reward spread, near-degenerate groups, and mean alignment.

\subsubsection{Per-Seed Main Results}
Table~\ref{tab:bright-current-seeds} reports the individual runs underlying
Table~\ref{tab:main-posttrain}. With GPT-4 reasoning queries, RELER exceeds
InfoNCE at every paired seed on all three backbones. Group means and SDs
are computed before rounding the displayed per-seed values.

\begin{table}[!htbp]
\centering
\caption{Per-seed BRIGHT nDCG@10 ($\times100$), averaged over 12 subsets.
The pretrained encoders are evaluated once. GPT-4 reasoning queries use
the released \texttt{gpt4\_reason.query} inputs.}
\label{tab:bright-current-seeds}
\tablesetup
\begin{tabular}{@{}llrrrrrr@{}}
\toprule
 & & \multicolumn{3}{c}{\tablehead{Original query}} & \multicolumn{3}{c}{\tablehead{GPT-4 reasoning query}} \\
\cmidrule(lr){3-5}\cmidrule(l){6-8}
\tablehead{Model} & \tablehead{Method} & 42 & 3407 & 2026 & 42 & 3407 & 2026 \\
\midrule
Q3E-0.6B & Q3E-0.6B & 15.10 & -- & -- & 20.38 & -- & -- \\
 & InfoNCE & 22.50 & 22.56 & 22.87 & 26.40 & 26.59 & 26.56 \\
 & LambdaLoss & 19.13 & 19.30 & 19.67 & 23.61 & 23.96 & 24.34 \\
 & RELER & 23.27 & 23.57 & 23.29 & 27.15 & 27.19 & 27.10 \\
\midrule
Q3E-4B & Q3E-4B & 18.70 & -- & -- & 22.01 & -- & -- \\
 & InfoNCE & 30.17 & 30.17 & 29.89 & 30.12 & 30.70 & 30.05 \\
 & LambdaLoss & 25.44 & 24.76 & 24.85 & 25.96 & 25.49 & 25.61 \\
 & RELER & 29.98 & 31.06 & 30.33 & 30.77 & 31.52 & 30.92 \\
\midrule
BGE-M3 & BGE-M3 & 10.66 & -- & -- & 21.45 & -- & -- \\
 & InfoNCE & 12.67 & 12.51 & 12.71 & 20.69 & 20.63 & 20.52 \\
 & LambdaLoss & 11.38 & 11.33 & 11.49 & 19.12 & 19.03 & 19.24 \\
 & RELER & 14.48 & 14.46 & 14.54 & 23.77 & 23.91 & 23.84 \\
\bottomrule
\end{tabular}

\end{table}

\FloatBarrier

\subsection{BRIGHT Reference Baselines}\label{sec:bright-reference-baselines}
Table~\ref{tab:published-bright-references} shows original-query BRIGHT
results from Table~2 of \citet{chen2026reasonembed}, excluding their MS MARCO
fine-tuning controls and pretrained Qwen3-Embedding rows, and including
their ReasonEmbed-Qwen3-4B Redapter model.
It also includes our pretrained 4B baseline and trained recipes from
Table~\ref{tab:main-posttrain}. Published results were not rerun under our
evaluation protocol; their scores provide context rather than controlled
comparisons with RELER.

\begin{table}[!htbp]
\centering
\caption{Original-query BRIGHT nDCG@10 ($\times100$). Literature rows are
from Table~2 of \citet{chen2026reasonembed}; MS MARCO fine-tuning controls
and pretrained Qwen3-Embedding rows are omitted. ReasonEmbed-Qwen3-4B uses Redapter.
Our 4B rows repeat Table~\ref{tab:main-posttrain}; Avg. reports three-seed
mean $\pm$ sample SD for trained recipes. Literature and our results use
different training and evaluation protocols. Abbreviations follow the main table.}
\label{tab:published-bright-references}
\begingroup
\tablesetup
\setlength{\tabcolsep}{2.2pt}
\begin{tabular}{@{}l*{7}{r}@{\hspace{6pt}}rr@{\hspace{6pt}}rrr@{\hspace{7pt}}l@{}}
\toprule
 & \multicolumn{7}{c}{StackExchange} & \multicolumn{2}{c}{Coding} & \multicolumn{3}{c}{Theorem-based} & \\
\cmidrule(lr){2-8}\cmidrule(lr){9-10}\cmidrule(lr){11-13}
Method & Bio. & Earth. & Econ. & Psy. & Rob. & Stack. & Sus. & Leet. & Pony & AoPS & TheoQ. & TheoT. & Avg. \\
\midrule
\multicolumn{14}{@{}l}{\textit{General-purpose methods}} \\
BM25 & 18.9 & 27.2 & 14.9 & 12.5 & 13.6 & 18.4 & 15.0 & 24.4 & 7.9 & 6.2 & 10.4 & 4.9 & 14.5 \\
OpenAI-3-Large & 23.3 & 26.7 & 19.5 & 27.6 & 12.8 & 14.3 & 20.5 & 23.6 & 2.4 & 8.5 & 23.5 & 11.7 & 17.9 \\
Google-Gecko-1B-768 & 22.7 & 34.8 & 19.6 & 27.8 & 15.7 & 20.1 & 17.1 & 29.6 & 3.6 & 9.3 & 23.8 & 15.9 & 20.0 \\
GritLM-7B & 24.8 & 32.3 & 18.9 & 19.8 & 17.1 & 13.6 & 17.8 & 29.9 & 22.0 & 8.8 & 25.2 & 21.2 & 21.0 \\
gte-Qwen2-7B-instruct & 34.1 & 42.6 & 18.2 & 27.4 & 13.2 & 17.3 & 20.9 & 30.4 & 2.2 & 13.3 & 30.6 & 32.6 & 23.5 \\
\midrule
\multicolumn{14}{@{}l}{\textit{Tailored methods for reasoning-intensive retrieval}} \\
ReasonIR-8B & 26.2 & 31.4 & 23.3 & 30.0 & 18.0 & 23.9 & 20.5 & 35.0 & 10.5 & 14.7 & 31.9 & 27.2 & 24.4 \\
RaDeR-gte-Qwen2-7B & 34.6 & 38.9 & 22.1 & 33.0 & 14.8 & 22.5 & 23.7 & 37.3 & 5.0 & 10.2 & 28.4 & 35.1 & 25.5 \\
Seed-1.5-Embedding & 34.8 & 46.9 & 23.4 & 31.6 & 19.1 & 25.4 & 21.0 & 43.2 & 4.9 & 12.2 & 33.3 & 30.5 & 27.2 \\
DIVER-Retriever (4B) & 41.8 & 43.7 & 21.7 & 35.3 & 21.0 & 21.2 & 25.1 & 37.6 & 13.2 & 10.7 & 38.4 & 37.3 & 28.9 \\
\shortstack[l]{ReasonEmbed-Q3E-4B\\(Redapter)} & 55.4 & 54.5 & 34.9 & 46.9 & 34.0 & 36.1 & 37.4 & 34.5 & 13.6 & 11.3 & 41.4 & 45.1 & 37.1 \\
\midrule
\multicolumn{14}{@{}l}{\textit{Our Q3E-4B adaptation}} \\
Q3E-4B & 17.2 & 34.6 & 16.4 & 23.0 & 13.0 & 15.9 & 17.1 & 21.2 & 2.0 & 3.7 & 19.6 & 40.9 & 18.70 \\
LambdaLoss & 32.3 & 35.5 & 27.5 & 36.0 & 19.9 & 27.0 & 30.8 & 18.5 & 3.6 & 3.8 & 23.3 & 42.0 & \tablescore{25.01}{0.37} \\
InfoNCE & 48.5 & 47.0 & 32.0 & 42.3 & 26.7 & 34.7 & 35.9 & 15.2 & 6.6 & 3.8 & 23.5 & 44.7 & \tablescore{30.08}{0.16} \\
\rowcolor{relerrow}
RELER & 48.9 & 50.2 & 32.9 & 44.3 & 25.8 & 33.4 & 33.6 & 13.8 & 11.3 & 3.6 & 20.7 & 46.9 & \tablescore{30.46}{0.55} \\
\bottomrule
\end{tabular}
\endgroup

\end{table}

\FloatBarrier

\subsection{Complete BRIGHT Ablation Results}
\label{sec:retrieval-breakdown}
\label{sec:additional-ablations}

The BRIGHT ablation and sensitivity tables use Q3E-0.6B with original queries,
$G=64$, 113 training steps, and seeds 42, 3407, and 2026.
Subset entries are three-seed means of nDCG@10 ($\times100$);
Avg. is the 12-subset macro-average with sample SD across seeds.
Subset means are computed from the rounded per-subset exports; Avg. and SD
retain the grouped export precision. Bold marks the highest displayed mean
within each block, and shading marks the main RELER setting.

Table~\ref{tab:bright-ablation-subsets} gives the 12-subset breakdown of
Table~\ref{tab:core-ablations}, retaining its row labels and order.
All rows use own-query candidates and $\rho=0.70$.
The first block compares full RELER with its unprojected RLOO control,
disabling CMP in both reward components while retaining the other settings.
The second compares CMP, rollouts, and one-sided policies under standalone graded nDCG;
the third reports binary rewards with product rollouts and RLOO+CMP.
Paired rollouts use one-to-one query--bundle pairings, whereas product
rollouts evaluate all $G^2$ pairings. The one-sided policies sample only
the named embeddings while updating the shared encoder. These rollout
comparisons all use graded nDCG@10 on the own-query candidate set.

\begin{table}[!htbp]
\centering
\caption{Complete subset results for Table~\ref{tab:core-ablations}.
RELER uses $\lambda=0.5$. Subset abbreviations
follow Table~\ref{tab:main-posttrain}.}
\label{tab:bright-ablation-subsets}
\begingroup
\compacttablesetup
\setlength{\tabcolsep}{1.9pt}
\begin{tabular}{@{}l*{12}{r}l@{}}
\toprule
 & \multicolumn{7}{c}{\tablehead{StackExchange}} & \multicolumn{2}{c}{\tablehead{Coding}} & \multicolumn{3}{c}{\tablehead{Theorem-based}} & \\
\cmidrule(lr){2-8}\cmidrule(lr){9-10}\cmidrule(lr){11-13}
\tablehead{Training variant} & \tablehead{Bio.} & \tablehead{Earth.} & \tablehead{Econ.} & \tablehead{Psy.} & \tablehead{Rob.} & \tablehead{Stack.} & \tablehead{Sus.} & \tablehead{Leet.} & \tablehead{Pony} & \tablehead{AoPS} & \tablehead{TheoQ.} & \tablehead{TheoT.} & \tablehead{Avg.} \\
\midrule
\multicolumn{14}{l}{\textit{Graded nDCG + pairwise}} \\
\cmidrule{1-14}
\rowcolor{relerrow}
\textit{(a)} RELER & \textbf{45.9} & \textbf{37.3} & \textbf{28.6} & \textbf{34.0} & \textbf{18.4} & \textbf{29.3} & \textbf{25.2} & 9.4 & \textbf{2.2} & 2.3 & 19.0 & 28.9 & \tablescore{\textbf{23.38}}{0.17} \\
\textit{(b)} RELER w/o CMP & 33.6 & 35.3 & 27.1 & 32.0 & 18.0 & 27.0 & 23.7 & \textbf{10.2} & 1.3 & \textbf{2.7} & \textbf{19.7} & \textbf{32.4} & \tablescore{21.91}{0.70} \\
\midrule
\multicolumn{14}{l}{\textit{Graded nDCG}} \\
\cmidrule{1-14}
\textit{(c)} RLOO+CMP / product & \textbf{40.5} & \textbf{38.0} & \textbf{28.0} & \textbf{34.7} & \textbf{18.2} & \textbf{25.4} & \textbf{23.7} & 10.1 & \textbf{2.3} & 2.7 & 19.4 & 29.6 & \tablescore{\textbf{22.72}}{0.18} \\
\textit{(d)} RLOO / product & 27.7 & 31.9 & 27.3 & 32.4 & 17.3 & 23.8 & 23.1 & 10.2 & 1.3 & 3.0 & \textbf{19.7} & 32.4 & \tablescore{20.85}{0.68} \\
\textit{(e)} RLOO / paired & 20.6 & 29.4 & 24.6 & 27.8 & 14.2 & 20.6 & 20.7 & 12.2 & 1.2 & \textbf{3.4} & 18.5 & 31.4 & \tablescore{18.71}{0.53} \\
\textit{(f)} RLOO / query only & 15.5 & 21.3 & 19.2 & 20.0 & 14.2 & 16.5 & 18.8 & \textbf{16.2} & 1.1 & 2.8 & 18.0 & 26.7 & \tablescore{15.85}{0.26} \\
\textit{(g)} RLOO / document only & 21.0 & 32.0 & 24.6 & 25.7 & 14.4 & 18.7 & 18.1 & 10.8 & 1.1 & 2.5 & 17.5 & \textbf{34.8} & \tablescore{18.45}{1.56} \\
\midrule
\multicolumn{14}{l}{\textit{Binary rewards (product rollouts, RLOO+CMP)}} \\
\cmidrule{1-14}
\textit{(h)} Binary nDCG & 27.0 & 24.1 & 21.3 & 21.2 & 12.2 & \textbf{29.2} & \textbf{17.1} & 7.5 & \textbf{1.4} & 2.0 & 16.1 & 30.1 & \tablescore{17.45}{0.65} \\
\textit{(i)} Binary MRR & \textbf{32.9} & \textbf{28.0} & \textbf{22.9} & \textbf{22.2} & \textbf{13.8} & 27.7 & 16.5 & \textbf{8.2} & 1.2 & \textbf{2.8} & \textbf{16.7} & \textbf{34.0} & \tablescore{\textbf{18.90}}{0.43} \\
\bottomrule
\end{tabular}

\endgroup
\end{table}

\FloatBarrier

\subsection{Pairwise Weight and Alignment}
\label{sec:pairwise-ablations}

Table~\ref{tab:bright-sensitivity-subsets} gives the full subset results
behind Figure~\ref{fig:pairwise-weight}(c,d). Both sweeps use own-query
candidates and product rollouts with RLOO+CMP. The first varies $\lambda$ at
$\rho=0.70$; the second varies $\rho$ at $\lambda=0.5$.
The main setting ($\lambda=0.5$, $\rho=0.70$) is repeated in both blocks.
The gain from $\lambda=0.5$ comes from StackExchange; Coding and
Theorem-based retrieval score lower than with $\lambda=0$.
Changing alignment has a smaller effect on the average. Although
$\rho=0.80$ scores 0.08 points above the main setting, it does not
improve every paired seed.

\begin{table}[!htbp]
\centering
\caption{Complete BRIGHT subset results for the pairwise-weight and
alignment sweeps in Figure~\ref{fig:pairwise-weight}(c,d). The first column
is $\lambda$ in the upper block and $\rho$ in the lower block.}
\label{tab:bright-sensitivity-subsets}
\begingroup
\compacttablesetup
\setlength{\tabcolsep}{2.6pt}
\begin{tabular}{@{}l*{12}{r}l@{}}
\toprule
 & \multicolumn{7}{c}{\tablehead{StackExchange}} & \multicolumn{2}{c}{\tablehead{Coding}} & \multicolumn{3}{c}{\tablehead{Theorem-based}} & \\
\cmidrule(lr){2-8}\cmidrule(lr){9-10}\cmidrule(lr){11-13}
\tablehead{Setting} & \tablehead{Bio.} & \tablehead{Earth.} & \tablehead{Econ.} & \tablehead{Psy.} & \tablehead{Rob.} & \tablehead{Stack.} & \tablehead{Sus.} & \tablehead{Leet.} & \tablehead{Pony} & \tablehead{AoPS} & \tablehead{TheoQ.} & \tablehead{TheoT.} & \tablehead{Avg.} \\
\midrule
\multicolumn{14}{l}{\textit{Pairwise weight $\lambda$ ($\rho=0.70$)}} \\
\cmidrule{1-14}
0 & 40.5 & \textbf{38.0} & 28.0 & \textbf{34.7} & 18.2 & 25.4 & 23.7 & \textbf{10.1} & \textbf{2.3} & \textbf{2.7} & \textbf{19.4} & \textbf{29.6} & \tablescore{22.72}{0.18} \\
\rowcolor{relerrow}
0.5 & \textbf{45.9} & 37.3 & \textbf{28.6} & 34.0 & \textbf{18.4} & 29.3 & \textbf{25.2} & 9.4 & 2.2 & 2.3 & 19.0 & 28.9 & \tablescore{\textbf{23.38}}{0.17} \\
1.0 & 43.6 & 34.9 & 27.2 & 32.8 & 18.1 & \textbf{29.7} & 23.9 & 9.0 & 2.0 & 1.9 & 18.9 & 28.9 & \tablescore{22.57}{0.18} \\
\midrule
\multicolumn{14}{l}{\textit{Alignment $\rho$ ($\lambda=0.5$)}} \\
\cmidrule{1-14}
0.60 & \textbf{46.2} & \textbf{38.3} & \textbf{28.8} & \textbf{34.3} & 18.2 & \textbf{29.6} & 24.9 & 8.8 & 2.1 & \textbf{2.7} & 18.9 & 27.5 & \tablescore{23.36}{0.25} \\
\rowcolor{relerrow}
0.70 & 45.9 & 37.3 & 28.6 & 34.0 & 18.4 & 29.3 & \textbf{25.2} & 9.4 & \textbf{2.2} & 2.3 & 19.0 & 28.9 & \tablescore{23.38}{0.17} \\
0.80 & 45.1 & 37.5 & 28.6 & \textbf{34.3} & \textbf{19.6} & 28.7 & 24.9 & 10.0 & 2.0 & 2.2 & 19.0 & 29.6 & \tablescore{\textbf{23.46}}{0.25} \\
0.90 & 42.4 & 37.1 & 28.2 & 33.8 & 19.1 & 28.2 & 24.6 & \textbf{10.3} & 2.0 & 2.2 & \textbf{19.5} & \textbf{30.6} & \tablescore{23.17}{0.19} \\
\bottomrule
\end{tabular}

\endgroup
\end{table}

\FloatBarrier

\clearpage
\subsection{Gradient Variance: Protocol and Results}
\label{sec:gradient-probe-protocol}

\paragraph{Matched main-recipe probes.}
We fix Q3E-0.6B before fine-tuning and at steps 50 and 113 of the
seed-3407 RELER run ($\lambda=0.5$). At each state, three fixed training
microbatches (Stack Overflow, math-theorem, and biology) receive 64
independent rollout draws. Both estimators use the same actions, own-query
candidate shortlist, graded reward table, and pairwise inputs. The
unprojected control replaces CMP in both reward components with RLOO
score-function terms; it does not leave the pairwise CMP path active.
Draw-level hashes and exact pairwise-reward agreement check the pairing.
Dropout is disabled and no optimizer update is made. The measurements
include every trainable encoder parameter.

For gradients $g_1,\ldots,g_N$ in one fixed state and microbatch, define
$\bar g=N^{-1}\sum_k g_k$ and
$\widehat V=(N-1)^{-1}\sum_k\|g_k-\bar g\|^2$, with $N=64$.
The bias-corrected squared signal estimate is
$\widehat S^2=\|\bar g\|^2-\widehat V/N$, and we report the single-draw
signal-to-noise ratio $\sqrt{\max(\widehat S^2,0)/\widehat V}$.
For paired differences
$\delta_k=g_k^{\rm CMP}-g_k^{\rm RLOO}$, we compare
$\|\bar\delta\|^2-\widehat V_\delta/N$ with zero as a descriptive
Monte Carlo resolution check, not a formal significance test.
\paragraph{Complete-gradient measurements.}
The variance reduction holds in all nine probes
(Table~\ref{tab:gradient-probes}): CMP lowers full-gradient variance by
291--590 times, and raises the single-draw SNR above one throughout.
The cost is an 11--14\% increase in forward/backward time per draw, so
the reduction remains large when measured by variance times runtime.
These are fixed-state measurements of rollout noise; the BRIGHT seed SD
also reflects the subsequent optimization trajectory.

\begin{table}[htbp]
\centering
\caption{Complete-encoder gradient measurements for the combined reward.
Each row uses 64 paired rollout draws at one fixed encoder state and
microbatch. $V$ is the sample gradient variance and $t$ is mean
forward/backward time per draw. Ratios compare CMP to RLOO; SNR uses the
bias-corrected signal estimate.}
\label{tab:gradient-probes}
\tablesetup
\begin{tabular}{@{}llrrrrr@{}}
\toprule
 & & \multicolumn{2}{c}{\tablehead{SNR}} & \multicolumn{3}{c}{\tablehead{CMP / RLOO}} \\
\cmidrule(lr){3-4}\cmidrule(l){5-7}
\tablehead{Encoder state} & \tablehead{Microbatch} & RLOO & CMP & $V$ & $t$ & $Vt$ \\
\midrule
Before fine-tuning & Stack Overflow & 0.25 & 5.35 & 0.00226 & 1.14 & 0.00258 \\
 & Math-theorem & 0.27 & 6.00 & 0.00182 & 1.11 & 0.00203 \\
 & Biology & 0.20 & 4.85 & 0.00169 & 1.13 & 0.00191 \\
\midrule
Step 50 & Stack Overflow & 0.23 & 4.07 & 0.00322 & 1.14 & 0.00367 \\
 & Math-theorem & 0.22 & 3.81 & 0.00321 & 1.11 & 0.00358 \\
 & Biology & 0.23 & 4.53 & 0.00248 & 1.13 & 0.00280 \\
\midrule
Step 113 & Stack Overflow & 0.23 & 3.95 & 0.00343 & 1.14 & 0.00392 \\
 & Math-theorem & 0.19 & 3.30 & 0.00305 & 1.12 & 0.00341 \\
 & Biology & 0.19 & 3.64 & 0.00283 & 1.13 & 0.00318 \\
\bottomrule
\end{tabular}

\end{table}

After correcting for sampling noise, the estimated squared difference
between the two mean gradients is nonpositive in five probes. In the
remaining four, it is at most 2.81\% of CMP's estimated squared signal.
These finite-sample measurements supplement the unbiasedness proof in
Appendix~\ref{sec:cp-details}. The proof bounds each summand's variance;
Table~\ref{tab:gradient-probes} measures the variance after summing all
contributions to the encoder gradient.

\paragraph{Gradient-cloud visualization.}
Figure~\ref{fig:gradient-noise}(a) uses the math-theorem microbatch before
fine-tuning. We project each complete gradient onto eight fixed random
sign vectors (seed 20260925), shared by both estimators, then display the
axes with the largest and smallest absolute CMP mean coordinates. Both
axes use the same scale, the absolute CMP mean on the first axis, and its
sign is oriented so that the CMP mean has first coordinate one. The inset
zooms into the CMP cloud. This selected two-dimensional view illustrates
the paired draws; all quantitative variance and SNR comparisons use the
complete parameter gradients in Table~\ref{tab:gradient-probes}.

\clearpage
\section{Fixed-Index RAG Experiments}
\label{sec:rag-appendix}

\subsection{Corpus and Training Data}

The fixed-index QA protocol in Section~\ref{sec:rag-results} builds a
Q3E-0.6B index over FlashRAG's \texttt{wiki18\_100w} release: 21,015,324
100-word passages from December 2018 English Wikipedia
\citep{karpukhin2020dense}.\footnote{Corpus:
\url{https://huggingface.co/datasets/RUC-NLPIR/FlashRAG_datasets}.}
The index and Qwen2.5-7B-Instruct generator stay frozen while the query
encoder adapts. All trained rows use NQ and HotpotQA
for one epoch, seed 42, and anchor coefficient 0.5. InfoNCE uses once-mined
candidates, whereas query-only RLOO searches the full index. The recipes
differ in both candidate access and objective. Training labels use DPR NQ
human-positive passages and HotpotQA supporting facts mapped to this corpus;
queries without complete evidence mappings are excluded.

\subsection{Training Objectives and Hyperparameters}
\label{sec:rag-training-details}

\paragraph{Passage judgments and rewards.}
The prepared NQ and HotpotQA records associate corpus passage IDs with
three judgments: external binary qrels, supporting-evidence membership,
and reference-answer containment. Qrels come from DPR human-positive NQ
passages and corpus-aligned HotpotQA supporting facts. For graded nDCG,
qrel positives receive grade 3, other supporting-evidence passages grade 2,
and remaining answer-containing passages grade 1; all others receive grade 0.
The highest-priority applicable grade is used. Gains are $2^\ell-1$.
Retrieved IDs are matched against these fixed passage judgments; IDs absent
from the judgment set receive zero gain. Ideal DCG is computed from the
highest grades in the query's complete prepared judgment set, independently
of the sampled retrieved list. Zero ideal DCG gives zero reward.
Answer-F1 rewards use the maximum normalized token F1 over reference
answers; the mixed reward averages this value and graded nDCG@10.

\paragraph{InfoNCE candidates and masking.}
The offline pool is mined once with the pretrained query encoder at depth
1,000, retaining all mapped qrel positives within this fixed-size pool. InfoNCE uses
the binary qrel positives in its numerator, at temperature 0.03.
Answer-containing passages that are not qrel positives are excluded from
the negative denominator, rather than treated as negatives. Query-only
RLOO instead searches the frozen full-corpus index for each sampled action.
All four trained recipes use the cosine anchor in
Appendix~\ref{sec:fixed-index-anchor}; Table~\ref{tab:rag-hyperparams}
lists the common optimizer and rollout settings.

\begin{table}[htbp]
\centering
\caption{Fixed-index RAG training and generation settings. The query encoder
is fully fine-tuned; the document index and generator remain frozen.}
\label{tab:rag-hyperparams}
\tablesetup
\begin{tabular}{ll}
\toprule
Setting & Value \\
\midrule
Training budget / seed & One epoch / 42 \\
Optimizer / learning rate & AdamW / $5\times10^{-6}$ \\
Global / per-device batch & 128 / 16 \\
Schedule / warm-up / weight decay & Cosine / 0.03 / 0.01 \\
Training query limit & 128 tokens \\
InfoNCE temperature / anchor coefficient & 0.03 / 0.5 \\
RL group size / alignment & 32 / 0.80 \\
RL baseline / advantage normalization & Leave-one-out / none \\
Retrieval reward / generator context depth & Top 10 / top 10 \\
Generator & Qwen2.5-7B-Instruct \\
Decoding / input limit / output limit & Greedy / 2,048 / 32 tokens \\
\bottomrule
\end{tabular}
\end{table}

\FloatBarrier

\clearpage
\subsection{Fixed-Index Query Anchor}
\label{sec:fixed-index-anchor}

Let $h_0(q)$ denote the normalized query embedding produced by the pretrained
encoder, and let $h_\theta(q)$ be the current deterministic query
embedding. For a training batch $\mathcal B$, the fixed-index experiments add
the following term to either the InfoNCE or query-only RLOO loss:
\begin{equation}
 \mathcal L_{\rm anchor}(\theta)
 =\alpha_{\rm anc}\frac{1}{|\mathcal B|}
 \sum_{q\in\mathcal B}\left(1-h_\theta(q)^\top h_0(q)\right),
 \qquad \alpha_{\rm anc}=0.5.
 \label{eq:fixed-index-anchor}
\end{equation}
The initial embeddings are computed once, before the first optimizer step,
using the same query prompt, tokenizer, pooling, and normalization as the
trainable encoder. They are stored by training-row position in FP16 and
treated as constants; gradients pass only through $h_\theta(q)$. The term
changes neither the retrieval or answer reward nor the frozen document index
and generator. Inference uses $h_\theta(q)$ directly and requires no anchor
lookup. Thus the penalty directly constrains training queries; any effect on
unseen queries comes through the learned encoder parameters.

The directional penalty also has a policy interpretation when both query vMF
distributions use the same fixed concentration $\kappa$. Writing
$A_d(\kappa)$ for the vMF mean resultant length, their KL divergence is
\begin{equation}
 D_{\rm KL}\!\left(
  \operatorname{vMF}(h_\theta(q),\kappa)
  \middle\|\operatorname{vMF}(h_0(q),\kappa)
 \right)
 =\kappa A_d(\kappa)\left(1-h_\theta(q)^\top h_0(q)\right).
 \label{eq:fixed-index-anchor-kl}
\end{equation}
Hence Eq.~\eqref{eq:fixed-index-anchor} is proportional to a KL penalty for
fixed $\kappa$; $\alpha_{\rm anc}$ is the coefficient on cosine distance,
not the coefficient on the KL itself. This equivalence does not apply if the
two policies use different concentrations.

\subsection{Generation and Evaluation}
We present passages in retrieval order, with numbered document blocks and
their titles, followed by the original question, using the generator's
chat template. The system prompt is: \emph{``Answer the question based on
the given documents. Only give me the answer and do not output any other
words.''} Prompts exceeding the input budget drop the lowest-ranked
passages first; if one remaining passage is still too long, its prefix is
retained together with the question. Greedy decoding uses temperature zero.
Evaluation caches generated answers by query, ordered passage IDs, context,
and generation configuration. EM/F1 lower-case text, remove punctuation
and English articles, and normalize whitespace; F1 uses token overlap and
requires an exact normalized match for yes/no/noanswer references or
predictions. Hit@10 checks normalized reference-answer containment in any
of the ten passages, independently of generator input truncation.

\subsection{Per-Dataset Results}
\label{sec:rag-details}

The averages in Table~\ref{tab:rag-main} conceal differences across datasets.
InfoNCE remains best on NQ, reaching 47.98 F1 despite its lower overall
average. Among the RL rewards, answer F1 exceeds graded nDCG in F1 on five
of seven datasets; NQ and Bamboogle are the exceptions.
The mixed reward adds 0.40 out-of-domain F1 points over graded nDCG,
alongside a 0.22-point Hit@10 gain. Its F1 margin over graded nDCG falls
from $+0.30$ points across seven datasets to $+0.13$ without Bamboogle.
Relative to answer-F1 training, the mixed reward's
$+0.24$-point seven-dataset F1 margin becomes $-0.25$ without Bamboogle.

Table~\ref{tab:rag-per-dataset} reports the dataset results and query counts
behind Table~\ref{tab:rag-main}. Each dataset has equal weight in the group
means despite the different query counts.

\begin{table}[!htbp]
\centering
\caption{Per-dataset fixed-index QA results ($\times100$), seed 42.
$n$ is the number of evaluation queries; 2Wiki denotes 2WikiMultihopQA.
The same document index, generator, and query counts are used for all recipes.}
\label{tab:rag-per-dataset}
\tablesetup
\begin{tabular}{@{}lrrrrrrr@{}}
\toprule
\tablehead{Method} & \tablehead{NQ} & \tablehead{HotpotQA} & \tablehead{PopQA} & \tablehead{TriviaQA} & \tablehead{2Wiki} & \tablehead{MuSiQue} & \tablehead{Bamboogle} \\
Queries ($n$) & 3,610 & 7,405 & 14,267 & 11,313 & 12,576 & 2,417 & 125 \\
\midrule
\multicolumn{8}{@{}l}{\textbf{Answer EM}} \\[2pt]
Q3E-0.6B & 34.99 & 32.59 & 42.22 & 61.00 & 28.76 & 7.36 & 19.20 \\
InfoNCE & 37.01 & 30.26 & 42.07 & 60.26 & 23.81 & 5.63 & 19.20 \\
RLOO, graded nDCG & 36.01 & 33.32 & 41.82 & 60.54 & 29.03 & 6.48 & 19.02 \\
RLOO, answer F1 & 35.94 & 33.49 & 42.40 & 61.43 & 28.98 & 7.27 & 16.80 \\
RLOO, nDCG + F1 & 36.12 & 33.52 & 42.70 & 61.20 & 28.75 & 6.80 & 21.60 \\
\midrule
\multicolumn{8}{@{}l}{\textbf{Answer F1}} \\[2pt]
Q3E-0.6B & 45.68 & 42.67 & 49.25 & 69.67 & 35.07 & 14.93 & 28.07 \\
InfoNCE & 47.98 & 40.17 & 49.20 & 69.08 & 30.12 & 12.43 & 27.97 \\
RLOO, graded nDCG & 46.69 & 43.48 & 48.96 & 69.29 & 35.16 & 14.26 & 27.73 \\
RLOO, answer F1 & 46.66 & 43.67 & 49.53 & 69.92 & 35.34 & 14.96 & 25.94 \\
RLOO, nDCG + F1 & 46.63 & 43.65 & 49.01 & 69.72 & 34.87 & 14.73 & 29.09 \\
\midrule
\multicolumn{8}{@{}l}{\textbf{Hit@10}} \\[2pt]
Q3E-0.6B & 76.84 & 54.50 & 76.74 & 79.24 & 46.25 & 27.89 & 29.60 \\
InfoNCE & 79.86 & 48.73 & 77.61 & 79.08 & 34.18 & 21.14 & 21.60 \\
RLOO, graded nDCG & 78.78 & 55.71 & 76.83 & 79.60 & 46.73 & 27.10 & 29.60 \\
RLOO, answer F1 & 77.61 & 55.93 & 76.90 & 79.53 & 46.14 & 29.29 & 31.20 \\
RLOO, nDCG + F1 & 78.09 & 55.65 & 76.72 & 79.76 & 46.26 & 27.60 & 30.60 \\
\bottomrule
\end{tabular}

\end{table}

\FloatBarrier

\FloatBarrier
\section{Embedding Training from a General-Purpose LLM}
\label{sec:initialization-study}

We also train embeddings directly from \texttt{Qwen3-0.6B}, a general-purpose
post-trained LLM, without embedding-specific initialization.
Both InfoNCE and RELER use the E2Rank training data \citep{liu2025e2rank},
full fine-tuning for 1,200 steps, learning rate $5\times10^{-6}$,
global batch size 128, and seed 42.
RELER uses graded nDCG@10 plus pairwise feedback ($\lambda=0.5$),
own-query candidates, product rollouts, RLOO+CMP, $G=64$, and $\rho=0.70$.
We evaluate the final checkpoints on MTEB English v2
\citep{muennighoff2023mteb,enevoldsen2025mmteb}, covering 41 tasks in seven types.

\begin{table}[!htbp]
\centering
\caption{Direct embedding training from \texttt{Qwen3-0.6B} on MTEB English v2
($\times100$), seed 42. Each type column averages its tasks;
Task avg. weights all 41 tasks equally, while Type avg. weights the seven
task-type means equally. Classif.: classification; Clust.: clustering;
Pair class.: pair classification; Rerank.: reranking; Summ.: summarization.
Bold marks the higher score in each column.}
\label{tab:initialization}
\tablesetup
\begin{tabular}{@{}l*{9}{r}@{}}
\toprule
\textbf{Objective} & \textbf{Classif.} & \textbf{Clust.} & \textbf{Pair class.} & \textbf{Rerank.} & \textbf{Retrieval} & \textbf{STS} & \textbf{Summ.} & \textbf{Task avg.} & \textbf{Type avg.} \\
\midrule
InfoNCE & 72.84 & 44.19 & \textbf{82.11} & 44.80 & \textbf{51.52} & 75.94 & \textbf{29.49} & 60.99 & \textbf{57.27} \\
\rowcolor{relerrow}
RELER & \textbf{73.41} & \textbf{45.81} & 79.76 & \textbf{45.13} & 50.04 & \textbf{76.81} & 26.61 & \textbf{61.01} & 56.80 \\
\bottomrule
\end{tabular}

\end{table}

RELER and InfoNCE are nearly tied in the all-task average
(Table~\ref{tab:initialization}), but InfoNCE leads in retrieval by 1.48
points and also has the higher task-type average. This experiment provides
no clear advantage for RELER
when starting from a general-purpose LLM. The comparison uses one seed
and complete recipes with different supervision and candidate handling.

\FloatBarrier

\clearpage
\section{Extended Related Work}
\label{sec:extended-related-work}

\paragraph{Retriever training and learning to rank.}
Contrastive training underlies dense retrieval and general-purpose embedding
adaptation \citep{karpukhin2020dense,wang2022text,behnamghader2024llm2vec}.
Learning-to-rank methods complement this representation-learning perspective
with pairwise and listwise surrogates, including differentiable approximations
to retrieval metrics \citep{burges2005learning,liu2009learning,
taylor2008softrank,pobrotyn2021neuralndcg}. Policy gradients have also been
used to optimize retrieval utility through MDP-based ranking policies and
stochastic rankings with fairness constraints
\citep{wei2017reinforcement,singh2019policy}.
RELER studies a one-step embedding action with offline relevance feedback.
Its contribution is the continuous policy over query and document
embeddings and its gradient estimator; matched InfoNCE and LambdaLoss
controls assess its value beyond supervised adaptation.

\paragraph{Reward-based retrieval and reasoning.}
ROPG-RL samples a profile document from a categorical policy over a fixed
candidate subset and uses a frozen LLM's generation metric as reward
\citep{salemi2024optimization}. RELER's fixed-index mode instead samples a
query vector and searches the full index, using retrieval or answer feedback
in our QA experiments. Retrieval-GRPO updates dense retrieval using
rewards on dynamically retrieved products \citep{liu2025retrievalgrpo},
while Rank-R1 and Search-R1 optimize generated reranking or search-and-reasoning
sequences \citep{zhuang2025rankr1,jin2025searchr1}.
GRACE optimizes rationale tokens with representation-based rewards and pools
rationale-conditioned states into embeddings \citep{sun2026grace}; Embed-RL
trains a separate reasoner with feedback from a frozen multimodal embedder
\citep{jiang2026embedrl}. The latter two approaches use generated reasoning to support
retrieval. RELER directly samples query and document embeddings during
training and retains deterministic single-vector inference.

\paragraph{Embedding policies and gradient estimation.}
vMF distributions also have precedents in hyperspherical representation
learning \citep{davidson2018hyperspherical}.
REINFORCE baselines, multi-sample LOO, and product-form recombination are
established estimation tools \citep{williams1992simple,greensmith2004variance,
kool2019buy,ahmadian2024back,kuntz2022product}. Their use in RELER's query and
document gradient estimator is detailed in Section~\ref{sec:policy-objective}
and Appendix~\ref{sec:reler-gradient}.
The closest antecedent of CMP is document-space projection:
\citet{wang2019variance} project exploration directions after interleaved
feedback in linear rankers. Our conditional-expectation construction includes
vMF means and accounts for each query--document reward cell and its LOO
baseline before updating a shared encoder. It applies Rao--Blackwell
averaging \citep{casella1996rao}, with a per-term variance guarantee;
variance of the summed encoder gradient is evaluated empirically.
The projection must retain every reward-relevant direction, so the guarantee
does not extend to using only returned full-corpus top-$k$ hits.

\end{document}